\documentclass[prd,showpacs,preprintnumbers,amsmath,amssymb]{revtex4}

\usepackage{graphics}
\usepackage{epsfig}
\usepackage{dcolumn}
\usepackage{bm}

\begin{document}
\title{$X(3872)$ and Other Possible Heavy Molecular States}
\author{Xiang Liu$^{1,2}$}\email{liuxiang@teor.fis.uc.pt}
\author{Zhi-Gang Luo$^1$}
\author{Yan-Rui Liu $^3$}
\author{Shi-Lin Zhu$^1$}
\email{zhusl@phy.pku.edu.cn}


\affiliation{$^1$Department of Physics, Peking University, Beijing
100871, China \\
$^2$Centro de F\'{i}sica Computacional, Departamento de
F\'{i}sica, Universidade de Coimbra,
P-3004-516, Coimbra, Portugal\\
$^3$Institute of High Energy Physics, P.O. Box 918-4, Beijing
100049, China}

\date{\today}

\begin{abstract}

We perform a systematic study of the possible molecular states
composed of a pair of heavy mesons such as $D\bar D$, $D^\ast\bar
D$, $D^\ast \bar D^\ast$ in the framework of the meson exchange
model. The exchanged mesons include the pseudoscalar, scalar and
vector mesons. Through our investigation, we find that (1) the
structure $X(3764)$ is not a molecular state; (2) There exists
strong attraction in the range $r < 1$ fm for the $D^*\bar D^*$
system with $J=0, 1$. If future experiments confirm $Z^+(4051)$ as
a loosely bound molecular state, its quantum number is probably
$J^{P}=0^+$. Its partner state $\Phi^{**0}$ may be searched for in
the $\pi^0\chi_{c1}$ channel; (3) The vector meson exchange
provides strong attraction in the $D^\ast \bar D$ channel together
with the pion exchange. A bound state solution exists with a
reasonable cutoff parameter $\Lambda\sim 1.4$ GeV. $X(3872)$ may
be accommodated as a molecular state dynamically although drawing
a very definite conclusion needs further investigation; (4) The
$B^\ast \bar B$ molecular state exists.

\end{abstract}


\maketitle

\section{Introduction}\label{sec1}

Since the observation of $Z^{+}(4430)$ \cite{Belle-4430}, the
Belle Collaboration reported two new resonance-like structures
$Z^+(4051)$ and $Z^+(4248)$ in the $\pi^+\chi_{c1}$ mass
distribution in the exclusive $\bar B^0\to K^-\pi^+\chi_{c1}$
decay. Their masses and widths are
$m_{Z^+(4051)}=(4051\pm14^{+20}_{-41})\, \mathrm{MeV}$,
$\Gamma_{Z^+(4051)}=(82^{+21+47}_{-17-22})\, \mathrm{MeV}$ and
$m_{Z^+(4248)}=(4248\pm14^{+44+180}_{-29-35})\,\mathrm{MeV}$,
$\Gamma_{Z^+(4248)}=(177^{+54+316}_{-39-61})\, \mathrm{MeV}$
\cite{Belle-new-two}. These charged hidden charm signals are good
candidates of either tetraquark states or heavy molecular states
if they are confirmed by future experiments.

Recently the BES Collaboration reported an anomalous line-shape
observed in the range of $3.650$ GeV to $3.872$ GeV by analyzing
the cross section of $e^+e^-\to \mathrm{hadrons}$. The anomalous
line-shape is composed of two possible enhancement structures
around 3.764 GeV and 3.779 GeV respectively \cite{BES}. The later
one is consistent with the well-established $\psi(3770)$ while the
mechanism of the first structure is not clear now, which is
denoted as $X(3764)$ in this work.

In the past five years, a series of observations of the
charmonium-like $X,\,Y,\,Z$ states, especially those states near
the threshold of two charmed mesons have stimulated the interest
in the possible existence of heavy molecular states greatly. The
presence of the heavy quarks lowers the kinetic energy while the
interaction between two light quarks could still provide strong
enough attraction.

In fact, Voloshin and Okun studied the interaction between a pair
of charmed mesons and proposed the possibilities of the molecular
states involving charmed quarks more than thirty years ago
\cite{Okun}. De Rujula, Geogi and Glashow speculated $\psi(4040)$
as a $D^*\bar{D}^*$ molecular state \cite{RGG}. T\"{o}rnqvist
studied the possible deuteron-like two-meson bound states such as
$D\bar{D}^*$ and $D^*\bar{D}^*$ using the quark-pion interaction
model \cite{Tornqvist}. Dubynskiy and Voloshin suggested the
possibility of the existence of a new resonance at the
$D^*\bar{D}^*$ threshold \cite{voloshin-1,voloshin}. Zhang,
Chiang, Shen and Zou studied the possible S-wave bound states of
two pseudoscalar mesons with the vector meson exchange in Ref.
\cite{zou}. The experimental observation of $X(3872)$ and
$Z^+(4430)$ motivated the extensive discussion of $X(3872)$ as a
$D\bar D^*$ molecular state
\cite{3872-Mole-1,3872-Mole-2,3872-Mole-3,3872-Mole-4,3872-Mole-5,
liu-3872,liu-4430-3872,YR-3872,Close-3872} and $Z^+(4430)$ as a
$D_1'D^*(D_1D^*)$ molecular state
\cite{rosner,Meng,liu-4430,liu-4430-1}.

It's interesting to note that the $Z^+(4051)$ enhancement
announced by the Belle collaboration and the $X(3764)$ signal
reported by the BES collaboration are near the thresholds of
$D^*\bar D^*$ and $D\bar D$ respectively. One may wonder whether
they could also be candidates of heavy molecular states. In this
work we perform a systematic study of three types of possible
heavy molecular states: the $D\bar D/B\bar B$ system (P-P), the
$D^\ast\bar D^\ast/B^\ast\bar B^\ast$ system (V-V), and the
$D^\ast\bar D/B^\ast\bar B$ system (P-V) using the formalism
developed in Ref. \cite{liu-3872,liu-4430,liu-4430-1}.

This paper is organized as follows. After the introduction, we
introduce some notations and present the flavor wave functions of
the S-wave P-P, V-V and P-V systems constructed by two heavy
flavor mesons. In Sec. \ref{sec3}, we collect the effective
Lagrangians in the derivation of the effective potential. Sections
\ref{secpp}-\ref{secpv} are for the P-P, V-V, P-V cases
respectively. The last section is the discussion and conclusion.

\section{Flavor wave functions of heavy molecular states}\label{sec2}

\begin{center}
\begin{figure}[htb]
\begin{tabular}{c}
\scalebox{0.5}{ \includegraphics{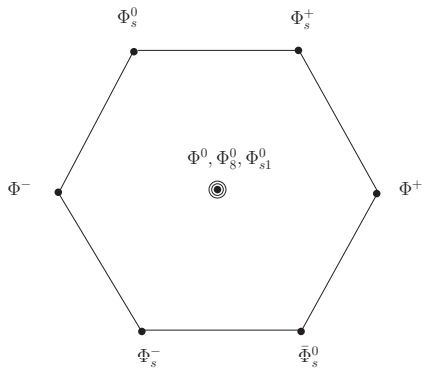}}
\end{tabular} \caption{The molecular multiplets composed of charmed and anit-charmed mesons. \label{state}}
\end{figure}
\end{center}

We study the possible molecular states composed of two
pseudoscalar (P-P) heavy mesons, two vector (V-V) heavy mesons,
one pseudoscalar and one vector (P-V) heavy mesons. The masses of
the $J^{P}=0^-,1^-$ heavy mesons are taken from PDG \cite{PDG} and
collected in Table \ref{mass}.
\begin{widetext}
\begin{center}
\begin{table}[htb]
\begin{tabular}{c||cccc|cccccc }\hline\hline
&\multicolumn{4}{c}{charmed-up(down) meson  }&&&&\multicolumn{2}{c}{charmed-strange meson  }&\\
\cline{2-10}

\raisebox{1.0ex}{$J^{P}$}&charged&mass (MeV)&neutral&mass (MeV)&
&&&charged&mass (MeV)\\\cline{1-10}

$0^-$&$D^{\pm}$&1869.3& $D^{0}$&1864.5&&&&$D_s^{\pm}$&1968.2
\\
$1^-$&$D^{*\pm}$&2010.0&$D^{*0}$&2006.7&&&&$D_s^{*\pm}$&2112.0
\\\hline\hline
&\multicolumn{4}{c}{bottom-up(down) meson  }&&&&\multicolumn{2}{c}{bottom-strange meson  }&\\
\cline{2-10}

\raisebox{1.0ex}{$J^{P}$}&charged&mass (MeV)&neutral&mass (MeV)&
&&&neutral&mass (MeV)\\\cline{1-10}

$0^-$&$B^{\pm}$&5279.1& $B^{0}$&5279.5&&&&$B_s^{0}$&5365.1
\\
$1^-$&$B^{*\pm}$&5325.1&$B^{*0}$&5325.1&&&&$B_s^{*0}$&5412.0
\\\hline\hline
\end{tabular}
\caption{The masses of heavy mesons in the H doublet \cite{PDG}.
\label{mass}}
\end{table}
\end{center}
\end{widetext}
The P-P type is categorized as two systems, i.e. $\mathcal{D-\bar
D}$, $\mathcal{\bar B-B}$. Here $\mathcal{D}$, $\mathcal{\bar D}$,
$\mathcal{B}$ and $\mathcal{\bar B}$ denote $(D^0,D^{+},D_s^+)$,
$(\bar D^0,D^{-},D_s^-)$, $(B^{+},B^0,B_s^0)$ and $(B^{-},\bar
B^0,\bar B_s^0)$ triplets respectively. In the following, we
illustrate their flavor wave functions with the $\mathcal{D-\bar
D}$ type as an example. Since charmed mesons belong to the
fundamental representation of flavor $SU(3)$, the $\mathcal{D-\bar
D}$ system form an octet and a singlet: $3\otimes \bar
3=8\oplus1$. The corresponding flavor wave functions are
$\Phi_{s}^{+}=\bar{D}^0 D_{s}^+$, $ \Phi^{+}=\bar{D}^0 D^+$,
$\Phi_{s}^{0}=D^-D_s^+$, $ \Phi^{0}=\frac{1}{\sqrt{2}}(\bar{D}^0
D^0-D^-D^+)$, $\bar{\Phi}_s^{0}={D}_s^- D^+$,
$\Phi_{8}=\frac{1}{\sqrt{6}}(\bar{D}^0D^0+D^-D^+ -2 D_s^- D_s^+)$,
$ \Phi^{-}={D}^- D^0$, $\Phi_{s}^{-}={D}_s^- D^0$, $
\Phi_1=\frac{1}{\sqrt{3}}(\bar{D}^0 D^0+D^-D^+ +D_{s}^- D_s^+)$.
Through ideal mixing, we have
$\Phi_8^0=\frac{1}{\sqrt{2}}(\bar{D}^0D^0+D^-D^+)$ and $
\Phi_{s1}^0=D_s^-D_s^+$, where $\Phi_{8}^0$ is an isoscalar while
$\Phi^0$ is an isovector. The above states are shown in Fig.
\ref{state} and collected in Table \ref{table} together with the
$\mathcal{\bar B-B}$ type states. The flavor wave functions of the
$D^\ast\bar D^\ast$ and $B^\ast\bar B^\ast$ systems are
constructed similarly, which are denoted as $\Phi^{**}$ and
$\Omega^{**}$ respectively.

\begin{widetext}
\begin{center}
\begin{ruledtabular}
\begin{table}[htb]
\begin{tabular}{c|c||c|cccccccc}
\multicolumn{2}{c}{ $\mathcal{D-\bar D}$}&\multicolumn{2}{c}{
$\mathcal{\bar B- B}$}
\\\hline State& wave function&State& wave function\\\hline
$\Phi_s^+$&$\bar{D}^0 D_s^+$&$\Omega_s^+$&$B^+\bar{B}_{s}^0$\\
$\Phi^+$&$\bar{D}^0D^+$&
$\Omega^+$&$B^+\bar B^0$\\
$\Phi_s^0$&$D^-D_s^+$&$\Omega_s^0$&${B}^0 \bar B_s^0$\\
$\Phi^0$&$\frac{1}{\sqrt{2}}(\bar{D}^0D^0-D^-D^+)$&$\Omega^0$&$\frac{1}{\sqrt{2}}(B^+B^--B^0\bar B^0)$\\
$\bar\Phi^0_s$&$D_s^-D^+$&$\bar\Omega_s^0$&${B}_s^0 \bar B^0$\\
$\Phi^-$&$D^-D^0$&$\Omega^-$&${B}^0 B^-$\\
$\Phi_s^-$&$D_s^- D^0$&$\Omega_s^-$&${B}_s^0 B^-$\\
$\Phi_8^0$&$\frac{1}{\sqrt{2}}(\bar{D}^0D^0+D^-D^+)$&$\Omega_8^0$&$\frac{1}{\sqrt{2}}(B^+B^-+B^0\bar B^0)$\\
$\Phi_{s1}^0$&$D_s^-D_s^+$&$\Omega_{s1}^0$&$B_s^0 \bar
B_s^0$\\
\end{tabular}
\caption{The flavor wave functions of the $\mathcal{D-\bar D}$ and
$\mathcal{B-\bar B}$ systems. \label{table}}
\end{table}
\end{ruledtabular}
\end{center}
\end{widetext}

We label the $\mathcal{D-\bar D^*}$ and $\mathcal{B-\bar B^*}$
systems as $\Phi^{*}$ and $\Omega^{*}$ respectively and list their
flavor wave function in Table \ref{table-flavor}. Here we need
distinguish the $C$ parity for $\Phi^{*0}$, $\Phi_{8}^{*0}$ and
$\Phi_{s1}^{*0}$ in the $\mathcal{D-\bar D^*}$ system and
$\Omega^{*0}$, $\Omega_{8}^{*0}$ and $\Omega_{s1}^{*0}$ in the
$\mathcal{B-\bar B^*}$ system. The parameter $c=\mp1$ corresponds
to $C=\pm1$ respectively as pointed out in Ref.
\cite{liu-3872,liu-4430}. In this work we label those states with
the negative charge parity with a hat such as $\hat{\Phi}^{*0}$,
$\hat{\Phi}_{8}^{*0}$, $\hat{\Phi}_{s1}^{*0}$,
$\hat{\Omega}^{*0}$, $\hat{\Omega}_{8}^{*0}$ and
$\hat{\Omega}_{s1}^{*0}$. The charge parity of X(3872) is
positive.

\begin{widetext}
\begin{center}
\begin{ruledtabular}
\begin{table}[htb]
\begin{tabular}{c|c||c|cccccccc}
\multicolumn{2}{c}{ $\mathcal{D-\bar D^*}$}&\multicolumn{2}{c}{
$\mathcal{ B- \bar B^*}$}
\\\hline State& wave function&State& wave function\\\hline
$\Phi_s^{*+}/\hat \Phi_s^{*+}$&$\frac{1}{\sqrt{2}}(\bar{D}^{*0}
D_s^+ +c\,\bar{D}^0 D_{s}^{*+})$&$\Omega_s^{*+}/\hat
\Omega_s^{*+}$
&$\frac{1}{\sqrt{2}}(B^{*+}\bar{B}_{s}^0+c\,B^{+}\bar{B}_{s}^{*0})$\\

$\Phi^{*+}/\hat \Phi^{*+}$&$\frac{1}{\sqrt{2}}(\bar{D}^{*0} D^+
+c\,\bar{D}^0 D^{*+})$&
$\Omega^{*+}/\hat \Omega^{*+}$&$\frac{1}{\sqrt{2}}(B^{*+}\bar{B}^0+c\,B^{+}\bar{B}^{*0})$\\

$\Phi_s^{*0}/\hat \Phi_s^{*0} $&$\frac{1}{\sqrt{2}}( {D}^{*-}
D_s^+ +c\,{D}^-
D_{s}^{*+})$&$\Omega_s^{*0}/\hat \Omega_s^{*0}$&$\frac{1}{\sqrt{2}}(B^{*0}\bar{B}_{s}^0+c\,B^{0}\bar{B}_{s}^{*0})$\\

$\Phi^{*0}/\hat{\Phi}^{*0}$&$\frac{1}{2}[(\bar{D}^{*0}D^0-D^{*-}D^+)+c\,(\bar{D}^0D^{*0}-D^-D^{*+})]$
&$\Omega^{*0}/\hat{\Omega}^{*0}$&$\frac{1}{2}[(B^{*+}B^--B^{*0}\bar B^0)+c\,(B^+B^{*-}-B^0\bar B^{*0})]$\\

$\bar\Phi^{*0}_s/\hat{\bar{\Phi}}^{*0}_s
$&$\frac{1}{\sqrt{2}}(D_s^{*-}D^++c\,D_s^{-}D^{*+})$
&$\bar\Omega_s^{*0}/\hat{\bar{\Omega}}_{s}^{*0}$&$\frac{1}{\sqrt{2}}({B}_s^{*0} \bar B^0 +c\,{B}_s^{0} \bar B^{*0})$\\

$\Phi^{*-}/\hat
\Phi^{*-}$&$\frac{1}{\sqrt{2}}(D^{*-}D^0+c\,D^-D^{*0})$
&$\Omega^{*-}/\hat \Omega^{*-}$&$\frac{1}{\sqrt{2}}({B}^{*0} B^-+c\,{B}^0 B^{*-})$\\

$\Phi_s^{*-}/\hat \Phi_s^{*-}$&$\frac{1}{\sqrt{2}}(D_s^{*-}
D^0+c\,D_s^- D^{*0})$
&$\Omega_s^{*-}/\hat \Omega_s^{*-}$&$\frac{1}{\sqrt{2}}({B}_s^{*0} B^-+c\,{B}_s^0 B^{*-})$\\

$\Phi_8^{*0}/\hat{\Phi}_8^{*0}$&$\frac{1}{2}[(\bar{D}^{*0}D^0+D^{*-}D^+)+c\,(\bar{D}^0D^{*0}+D^-D^{*+})]$
&$\Omega_8^{*0}/\hat{\Omega}_8^{*0}$&$\frac{1}{2}[(B^{*+}B^-+B^{*0}\bar B^0)+c\,(B^+B^{*-}+B^0\bar B^{*0})]$\\

$\Phi_{s1}^{*0}/\hat{\Phi}_{s1}^{*0}$&$\frac{1}{\sqrt{2}}(D_s^{*-}D_s^++c\,D_s^-D_s^{*+})$
&$\Omega_{s1}^{*0}/\hat{\Omega}_{s1}^{*0}$&$\frac{1}{\sqrt{2}}(B_s^0
\bar B_s^0+c\,B_s^0 \bar
B_s^{*0})$\\
\end{tabular}
\caption{The flavor wave functions of the $\mathcal{D-\bar D^*}$
and $\mathcal{B-\bar B^*}$ systems. \label{table-flavor}}
\end{table}
\end{ruledtabular}
\end{center}
\end{widetext}

\section{The Effective Lagrangian in the
derivation of the potential}\label{sec3}

In the derivation of the potential, we need the the effective
Lagrangians, which are constructed based on the chiral symmetry
and heavy quark symmetry: \cite{lagrangian-hl,Casalbuoni,Falk}
\begin{eqnarray}
\mathcal{L}&=&igTr[H_b\gamma_\mu \gamma_5 \mathcal{A}_{ba}^\mu
\bar{H}_a]+i\beta Tr[H_b v^\mu
(\mathcal{V}_\mu-\rho_\mu)_{ba}\bar{H}_a]\nonumber\\&&+i\lambda
Tr[H_b\sigma^{\mu\nu}F_{\mu\nu}(\rho)\bar{H}_a]+g_\sigma
Tr[H_a\sigma \bar {H}_a],\label{all-lagrangian}
\end{eqnarray}
where the field $H$ is defined in terms of the $(0^-, 1^-)$
doublet
\begin{eqnarray}
H_b&=&\frac{1+\not v}{2 }[P_{b}^{*\mu}\gamma_\mu+iP_b \gamma_5].
\end{eqnarray}
$A_{ab}^{\mu}$ is the axial vector field with definition
\begin{eqnarray}
A_{ab}^{\mu}=\frac{1}{2}(\xi^{\dag}\partial^{\mu}\xi-\xi\partial^{\mu}\xi^{\dag})_{ab}=
\frac{i}{f_{\pi}}\partial^{\mu}\mathbb{P}_{ab}+\cdots
\end{eqnarray}
with $\xi=\exp(i\mathbb{P}/f_{\pi})$ and $f_\pi=132$ MeV. The
octet pseudoscalar and nonet vector meson matrices are
\begin{eqnarray}
\mathbb{P}&=&\left(\begin{array}{ccc}
\frac{\pi^{0}}{\sqrt{2}}+\frac{\eta}{\sqrt{6}}&\pi^{+}&K^{+}\\
\pi^{-}&-\frac{\pi^{0}}{\sqrt{2}}+\frac{\eta}{\sqrt{6}}&
K^{0}\\
K^- &\bar{K}^{0}&-\frac{2\eta}{\sqrt{6}}
\end{array}\right),
\end{eqnarray}
\begin{eqnarray}
\mathbb{V}&=&\left(\begin{array}{ccc}
\frac{\rho^{0}}{\sqrt{2}}+\frac{\omega}{\sqrt{2}}&\rho^{+}&K^{*+}\\
\rho^{-}&-\frac{\rho^{0}}{\sqrt{2}}+\frac{\omega}{\sqrt{2}}&
K^{*0}\\
K^{*-} &\bar{K}^{*0}&\phi
\end{array}\right).\label{vector}
\end{eqnarray}
The effective interaction Lagrangians at the tree level from Eq.
(\ref{all-lagrangian}) are
\begin{eqnarray} \nonumber
\mathcal{L}_{\mathcal{DD}\mathbb{V}}&=&-ig_{\mathcal{DD}\mathbb{V}}\left(\mathcal{D}_{a}^{\dag}\partial_\mu
\mathcal{D}_{b}-\mathcal{D}_{b}\partial_\mu
D_{a}^{\dag}\right)(\mathbb{V}^\mu)_{ab},\label{la-1} \\ \nonumber
\mathcal{L}_{\mathcal{DD}\sigma }&=&-2\,m_{\mathcal{D}}\,g_\sigma
\mathcal{D}_{a}\mathcal{D}_{a}^{\dag}\sigma,\label{la-2}\\
\nonumber
\mathcal{L}_{\mathcal{D^*D^*}\mathbb{P}}&=&\frac{1}{2}g_{\mathcal{D^*D^*}\mathbb{P}}
\varepsilon_{\mu\nu\alpha\beta}\left(D_{a}^{*\mu}\partial^\alpha
\mathcal{D}_{b}^{*\beta\dag}
-\mathcal{D}_{b}^{*\beta\dag}\partial^\alpha
 \mathcal{D}_{a}^{*\mu}\right)\partial^\nu \mathbb{P}_{ab},
\label{la-3}\\ \nonumber \mathcal{L}_{\mathcal{D^*D^*}\mathbb{V}
}&=&ig_{\mathcal{D^*D^*}\mathbb{V}}\left(\mathcal{D}_{a}^{*\nu\dag}\partial^\mu
\mathcal{D}_{\nu,b}^{*}-\mathcal{D}_{\nu b}^{*}\partial^\mu
\mathcal{D}_{a}^{*\nu\dag}\right)(\mathbb{V}_\mu)_{ab}
+4if_{\mathcal{D^*D^*}\mathbb{V}}D_{\mu a}^{*\dag} D_{\nu
b}^{*}(\partial^\mu \mathbb{V}^\nu -\partial^\nu
\mathbb{V}^\mu)_{ab}, \label{la-4}\\ \nonumber
\mathcal{L}_{\mathcal{D^*D^*}\sigma
}&=&2\,m_{\mathcal{D}^*}\,g_\sigma \mathcal{D}_{a}^{*\alpha}
\mathcal{D}_{\alpha a}^{*\dag}\sigma,\label{la-5}\\ \nonumber
\mathcal{L}_{\mathcal{D^*D}\mathbb{P}}&=&-ig_{\mathcal{D^*D}\mathbb{P}}\left(
\mathcal{D}_{a}\mathcal{D}_{\mu b}^{*\dagger}-\mathcal{D}_{\mu
a}^{*}\mathcal{D}^{\dagger}_b
\right)\partial^{\mu}\mathbb{P}_{ab},\label{la-6}\\ \nonumber
\mathcal{L}_{\mathcal{D^*D}\mathbb{V}}&=&-2f_{{\mathcal{D}^{*}\mathcal{D}\mathbb{V}}}
\varepsilon_{\mu\nu\alpha\beta}\left(\partial^{\mu}\mathbb{V}^{\nu}\right)_{ab}\left[\left(\mathcal{D}_{a}^{\dagger}
{\partial}^{\alpha}\mathcal{D}^{*\beta}_{b}-{\partial}^{\alpha}\mathcal{D}_{a}^{\dagger}
\mathcal{D}^{*\beta}_{b}\right)-\left(\mathcal{D}_{a}^{*\beta\dagger}
{\partial}^{\alpha}\mathcal{D}_{b}-{\partial}^{\alpha}\mathcal{D}_{a}^{*\beta\dagger}
\mathcal{D}_{b}\right)\right],\label{la-7}
\end{eqnarray}
where $\mathcal{D^{(*)}}$=(($\bar{D}^{0})^{(*)}$, $(D^{-})^{(*)}$,
$(D_{s}^{-})^{(*)}$). The relevant coupling constants are
\begin{eqnarray}
g_{{\mathcal{D}^{*}\mathcal{D}^{*}\mathbb{P}}}&=&\frac{g_{\mathcal{D}^*\mathcal{D}\mathbb{P}}}{\sqrt{m_{\mathcal{D}}m_{\mathcal{D}^*}}}=\frac{2g}{f_{\pi}},
\;\;\;g_{{\mathcal{DD}\mathbb
V}}=g_{{\mathcal{D}^{*}\mathcal{D}^{*}\mathbb V}}=\frac{\beta
g_{{V}}}{\sqrt{2}},\nonumber\\
f_{{\mathcal{D}^{*}\mathcal{D}\mathbb{V}}}&=&\frac{f_{{\mathcal{D}^{*}\mathcal{D}^{*}\mathbb{V}}}}{m_{{\mathcal{D}^*}}}=\frac{\lambda
g_{{V}}}{\sqrt{2}},\;\;\;
g_{{V}}=\frac{m_{_{\rho}}}{f_{\pi}},\;\;\;g_{\sigma}=\frac{g_{\pi}}{2\sqrt{6}},\label{para}
\end{eqnarray}
where $g_{_{V}},\;\beta$ and $\lambda$ are parameters in the
effective chiral Lagrangian that describe the interaction of heavy
mesons with light vector mesons \cite{Casalbuoni}. Following  Ref.
\cite{Isola}, we take $g=0.59$, $\beta=0.9$ and $\lambda=0.56$
$\mathrm{GeV}^{-1}$. $g_{\pi}=3.73$ \cite{Falk}.

In this work, we adopt the same formalism developed in Refs.
\cite{liu-3872,liu-4430,liu-4430-1} to derive the effective
potential. We compute the amplitudes of the elastic scattering of
two heavy mesons using the above effective Lagrangians. In order
to account for the structure effect of every interaction vertex,
we introduce the monopole type form factor (FF)
\cite{Tornqvist,FF}
\begin{eqnarray}
F(q)=\frac{\Lambda^2-m^2}{\Lambda^2-q^2}\; .\label{FFF}
\end{eqnarray}
Here $\Lambda$ is the phenomenological parameter around 1 GeV, and
$q$ denotes the four-momentum of the exchanged meson. The FF also
plays the role of regularizing the potential by imposing a
short-distance cutoff to cure the singularity of the effective
potential. Then we impose the constraint that the initial states
and final states should have the same angular momentum. After
averaging the potentials obtained with the Breit approximation in
the momentum space, we finally perform Fourier transformation to
derive the potentials in the coordinate space. In the following
sections, we illustrate the effective potentials in detail for the
P-P, V-V and P-V systems.

\section{The $\mathcal{D}-\bar{\mathcal{D}}$ case}\label{secpp}

\subsection{The potential of the P-P system}

The quantum number of the S-wave $\mathcal{D-\bar D}$ system is
$J^P=0^+$ and the C parity of the neutral states is positive. In
this work we consider the possible molecular states bound by the
force from the light pseudoscalar meson, vector meson and scalar
exchange. Parity and angular momentum conservation forbid the
exchange of a pseudoscalar meson between the $\mathcal{D-\bar D}$
pair. No suitable meson exchange is allowed for the $\Phi_s^{\pm}$
and $\Phi_s^0(\bar\Phi_s^0)$ cases. For the other states, the
effective potentials in the momentum space from the vector meson
and $\sigma$ meson exchange are
\begin{eqnarray}
\mathcal{V}_{_{\mathbb{V}}}(\textbf{q})\left[I_1^{\mathbb{V}\mathcal{D}_1\mathcal{D}_1}
,I_2^{\mathbb{V}\mathcal{ D}_2 \mathcal{
D}_2},m_{\mathcal{D}_1},m_{\mathcal{ D}_2},m_\mathbb{V}\right]
&=&-g_{\mathcal{DD}\mathbb{V}}^2\,\,I_1^{\mathbb{V}\mathcal{D}_1\mathcal{D}_1}
I_2^{\mathbb{V}\mathcal{ D}_2\mathcal{ D}_2}\left
(\frac{1}{\textbf{q}^2+m_\mathbb{V}^2}+\frac{\textbf{q}^2}{4m_{\mathcal{D}_1}
m_{\mathcal{ D}_2} m_\mathbb{V}^2}\right
),\label{pp1}\\
\mathcal{V}_\sigma(\textbf{q})&=&- {g_\sigma^2}
\frac{1}{\textbf{q}^2+m_\sigma^2},\label{pp2}
\end{eqnarray}
where $m_{\mathcal{D}}$ and $m_{\mathbb{V}}$ denote the masses of
charmed meson and exchanged vector meson respectively.
$I_1^{\mathbb{V}\mathcal{D}_1\mathcal{D}_1}$ and
$I_2^{\mathbb{V}\mathcal{ D}_2\mathcal{ D}_2}$ arise from the
coefficients related to the exchanged meson, which can be read
from the nonet vector meson matrix in Eq. (\ref{vector}). After
making the Fourier transformation, the three independent
structures in Eqs. (\ref{pp1})-(\ref{pp2}) read as
\begin{eqnarray}
&&\quad\,\,\mathbf{q}^2\quad\,\,\longrightarrow X(\Lambda,m,r),\\
&&\frac{1}{\mathbf{q}^2+m^2}\longrightarrow Y(\Lambda,m,r),\\
&&\frac{\mathbf{q}^2}{\mathbf{q}^2+m^2}\longrightarrow
Z(\Lambda,m,r)\; ,
\end{eqnarray}
where
\begin{eqnarray}
Y[\Lambda,m,r]&=&\frac{1}{4\pi r}\left(e^{-mr}-e^{-\Lambda
r}\right)-\frac{\xi^2}{8\pi \Lambda}e^{-\Lambda r},\\
Z[\Lambda,m,r]&=&-\frac{1}{r^2}\frac{\partial}{\partial r}\left(r^2\frac{\partial}{\partial r}\right)Y[\Lambda,m,r],\\
X[\Lambda,m,r]&=&\left[-\frac{1}{r^2}\frac{\partial}{\partial
r}\left(r^2\frac{\partial}{\partial
r}\right)+m^2\right]Z[\Lambda,m,r]
\end{eqnarray}
with $\xi=\sqrt{\Lambda^2-m^2}$. We have adopted the monopole FF
in Eq. (\ref{FFF}) to regularize the potential in Eqs.
(\ref{pp1})-(\ref{pp2}). Now the effective potentials in the
coordinate space read
\begin{eqnarray}
\mathcal{V}_{_{\mathbb{V}}}(r)\left[I_1^{\mathbb{V}\mathcal{D}_1\mathcal{D}_1}
,I_2^{\mathbb{V}\mathcal{ D}_2\mathcal{
D}_2},m_{\mathcal{D}_1},m_{\mathcal{ D}_2},m_\mathbb{V}\right]
&=&-g_{\mathcal{DD}\mathbb{V}}^2I_1^{\mathbb{V}\mathcal{D}_1\mathcal{D}_1}
I_2^{\mathbb{V}\mathcal{ D}_2\mathcal{ D}_2}\left
(Y[\Lambda,m_{\mathbb{V}},r]+\frac{X[\Lambda,m_{\mathbb{V}},r]}{4m_{\mathcal{D}_1}
m_{\mathcal{ D}_2} m_\mathbb{V}^2}\right
),\label{ppr1}\\
\mathcal{V}_\sigma(r)&=&- {g_\sigma^2}
Y[\Lambda,m_{\sigma},r].\label{ppr2}
\end{eqnarray}
The exchange potential for $\Phi^\pm$ in coordinate space is
\begin{eqnarray}
\mathcal{V}(r)_{Total}^{\Phi^\pm}&=&\mathcal{V}_{\mathbb{V}}(r)
\left[\frac{1}{\sqrt{2}},-\frac{1}{\sqrt{2}},m_{D^0},m_{D^{+}},m_{\rho^0}\right]+\mathcal{V}_{\mathbb{V}}(r)
\left[\frac{1}{\sqrt{2}},\frac{1}{\sqrt{2}},m_{D^0},m_{D^{+}},m_{\omega}\right]+\mathcal{V}_{\sigma}(r)\nonumber
\\&\approx&-g_{\mathcal{DD}\mathbb{V}}^2\left[\frac{Y[\Lambda,m_{\omega},r]}{2}-\frac{Y[\Lambda,m_{\rho},r]}{2}
-\frac{X[\Lambda,m_{\rho},r]}{8m_{D}^2m_{\rho}^2}+\frac{X[\Lambda,m_{\omega},r]}{8m_{D}^2m_{\omega}^2}\right]
- {g_\sigma^2} Y[\Lambda,m_{\sigma},r].\label{ppr-1}
\end{eqnarray}
The symbol "$\approx$" means the $SU(2)$ symmetry is assumed. For
the $\Phi^0$ and $\Phi_8^0$ states, the exchange potentials are
\begin{eqnarray}
\mathcal{V}(r)^{\Phi^0(\Phi_8^0)}_{Total}&=& \frac{1}{2}\Bigg\{
\mathcal{V}_{\mathbb{V}}(r)
\left[\frac{1}{\sqrt{2}},\frac{1}{\sqrt{2}},m_{D^0},m_{D^{0}},m_{\rho^0}\right]
\mp2\mathcal{V}_{\mathbb{V}}(r)
\left[1,1,{m_{D^0}},m_{D^+},m_{\rho^\pm}\right]
\nonumber\\&&+\mathcal{V}_{\mathbb{V}}(r)
\left[-\frac{1}{\sqrt{2}},-\frac{1}{\sqrt{2}},m_{D^+},m_{D^{-}},m_{\rho^0}\right]+
\mathcal{V}_{\mathbb{V}}(r)
\left[\frac{1}{\sqrt{2}},\frac{1}{\sqrt{2}},m_{D^0},m_{D^{0}},m_{\omega}\right]\nonumber\\&&
+ \mathcal{V}_{\mathbb{V}}(r)
\left[\frac{1}{\sqrt{2}},\frac{1}{\sqrt{2}},m_{D^+},m_{D^{-}},m_{\omega}\right]
+ 2\mathcal{V}_{\sigma}(r)\Bigg\}\nonumber\\&\approx&-
\frac{g_{\mathcal{DD}\mathbb{V}}^2}{2}(1\mp2)\left
(Y[\Lambda,m_{\rho},r]+\frac{X[\Lambda,m_{\rho},r]}{4m_{{D}}^2
m_{\rho}^2}\right )-\frac{g_{\mathcal{DD}\mathbb{V}}^2}{2}\left
(Y[\Lambda,m_{\omega},r]+\frac{X[\Lambda,m_{\omega},r]}{4m_{{D}}^2
m_{\omega}^2}\right )\nonumber\\&&-{g_\sigma^2}
Y[\Lambda,m_{\sigma},r] .\label{ppr-3}
\end{eqnarray}
Clearly the potential of $\Phi_{0}$ is the same as that of
$\Phi^{\pm}$ due to the SU(2) symmetry. The effective potential of
$\Phi_{s1}^0$ is
\begin{eqnarray}
\mathcal{V}(r)_{Total}^{\Phi_{s1}^0}&=&\mathcal{V}_{\mathbb{V}}(r)
\left[1,1,m_{D_s},m_{D_s},m_{\phi}\right]\nonumber\\&=&
-g_{\mathcal{DD}\mathbb{V}}^2\left
(Y[\Lambda,m_{\phi},r]+\frac{X[\Lambda,m_{\phi},r]}{4m_{{D}_s}^2
m_{\phi}^2}\right ),\label{ppr-2}
\end{eqnarray}
where only the $\phi$ meson exchange is allowed.

\subsection{Numerical results for the P-P system}\label{result-pp}

The input parameters include the coupling constants in Eq.
(\ref{para}), the masses of heavy mesons listed in Table
\ref{mass}, and the exchanged meson masses $m_{\pi}=139.5$ MeV,
$m_{\rho}=775.5$ MeV, $m_{\sigma}=600$ MeV, $m_{\eta}=547.5$ MeV,
$m_{\omega}=782.7$ MeV and $m_{\phi}=1019.5$ MeV.

\begin{center}
\begin{figure}[htb]
\begin{tabular}{ccc}
\scalebox{0.56}{\includegraphics{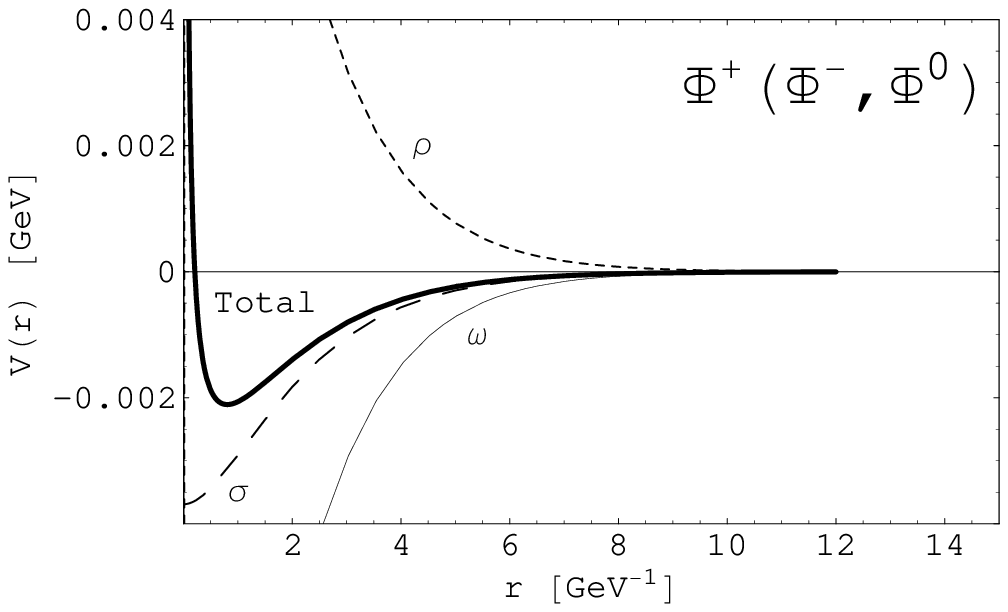}}&
\scalebox{0.56}{\includegraphics{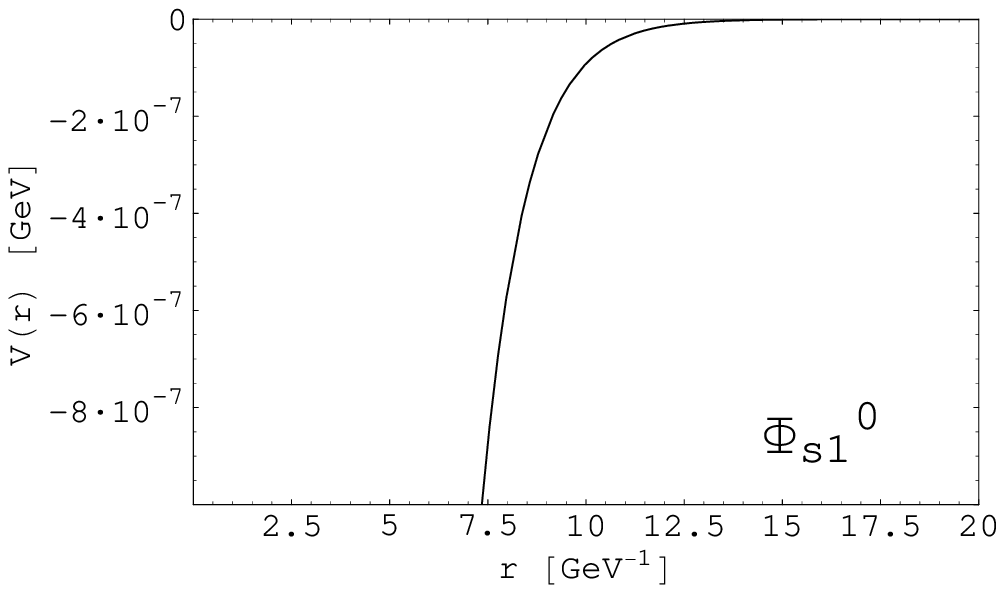}}&
\scalebox{0.56}{\includegraphics{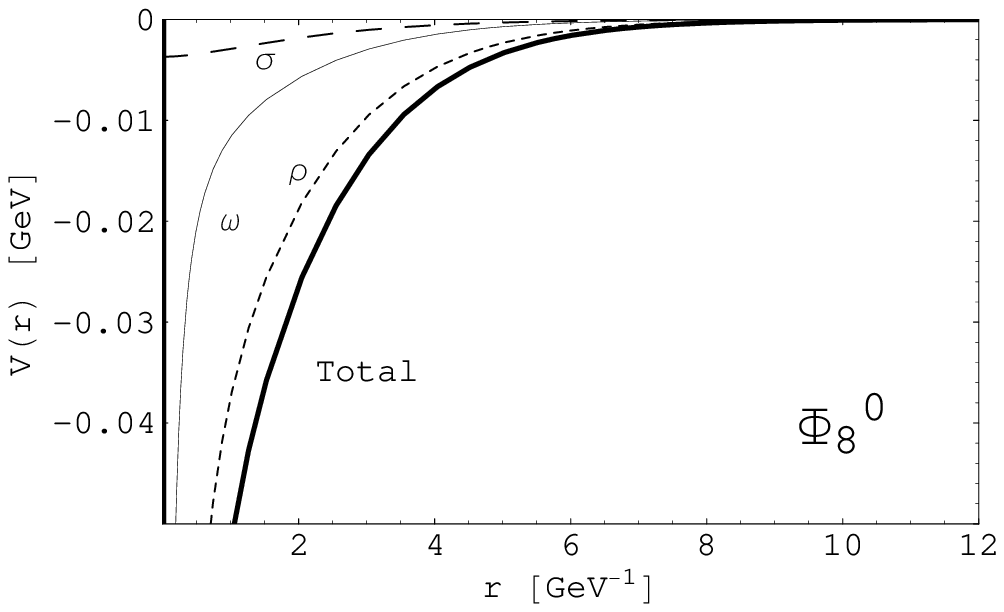}}\\(a)&(b)&(c)\\
\end{tabular}
\caption{Diagrams (a), (b) and (c) give respectively the variation
of potentials of $\Phi^{\pm}(\Phi)^{0}$, $\Phi_{s1}^0$ and
$\Phi_8^0$ states on $r$. \label{PP-potential}}
\end{figure}
\end{center}

We first plot the variation of the effective potentials of the
$\mathcal{D-\bar D}$ system with the cutoff $\Lambda=1$ GeV in
Fig. \ref{PP-potential}. For the $\Phi^{\pm}(\Phi^0)$ state, there
exist the $\rho$, $\omega$ and $\sigma$ meson exchanges. The
$\rho$ exchange force is repulsive while both the $\omega$ and
$\sigma$ meson exchange forces are attractive. The $\rho$ and
$\omega$ exchange forces cancel each other almost exactly. Thus
the total potential is attractive. For the state $\Phi_{s1}^0$,
the $\phi$ meson exchange potential is attractive. The exchange
potentials of $\rho$, $\omega$ and $\sigma$ mesons all are
attractive for $\Phi_8^0$. Thus its total effective potential is
attractive.

We adopt the MATSLISE package to solve the Schr\"{o}dinger
equation with the effective potentials of $\Phi^{\pm}(\Phi^0)$,
$\Phi_{s1}^0$ and $\Phi_8^0$ states. MATSLISE is a graphical
Matlab software package for the numerical study of regular
Sturm-Liouville problems, one-dimensional Schr\"{o}dinger
equations and radial Schr\"{o}dinger equations with a distorted
coulomb potential. It allows the fast and accurate computation of
the eigenvalues and the visualization of the corresponding
eigenfunctions \cite{matslise}.

\begin{center}
\begin{figure}[htb]
\begin{tabular}{c}
\scalebox{0.6}{ \includegraphics{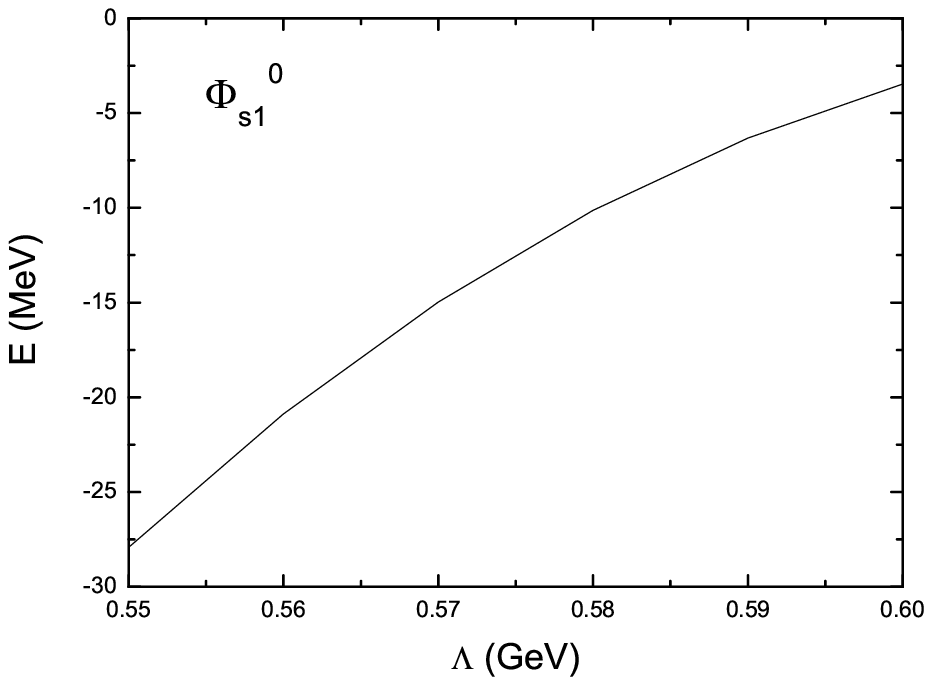}}
\end{tabular} \caption{The dependence of $E$ on $\Lambda$ for $\Phi_{s1}^0$. \label{PP-Phi_s1}}
\end{figure}
\end{center}

The binding energies with typical values of $\Lambda$ for
$\Phi^{\pm}(\Phi^0)$, $\Phi_{s1}^0$ and $\Phi_8^0$ are presented
in Table \ref{DD}, if there exists the solution by solving
Schr\"{o}dinger equation. $r_{\mathrm{rms}}$ denotes the
root-mean-square radius with the unit of fm. For
$\Phi^{\pm}(\Phi^0)$, we can not find bound state solutions in the
range of $\Lambda<10$ GeV, which indicates $\Phi^{\pm}(\Phi^0)$
does not exist. For both $\Phi_{s1}^0$ and $\Phi_8^0$, there exist
bound state solutions with $\Lambda$ around $0.5$ GeV. When the
binding energy $E$ becomes smaller, the $r_{\mathrm rms}$ becomes
larger. Similar observations hold for the $B-\bar B$ system. Now
the reduced masses of $\Omega_{s1}^0$ and $\Omega_8^0$ are
heavier. Hence the kinetic term is smaller. $\Omega_{s1}^0$ and
$\Omega_8^0$ states appear with a larger $\Lambda$ around 1.1 GeV.
As an example, we show the dependence of the binding energy of
$\Phi_{s1}^0$ on $\Lambda$ in Fig. \ref{PP-Phi_s1}. Moreover, we
increase and reduce both the vector coupling constant and scalar
coupling constant by a factor of two to see the variation of the
binding energy and the cutoff parameter. The numerical results are
collected in Tables \ref{DD-2-times} and \ref{DD-half-times}.

\begin{center}
\begin{ruledtabular}
\begin{table}[htb]
\begin{tabular}{c||ccccccccc}
&\multicolumn{3}{c}{$\mathcal{D-\bar D}$} \\\hline State&$\Lambda$
(GeV)&$E$ (MeV)& $r_{\mathrm{rms}}$ (fm)\\\hline
$\Phi^{\pm}(\Phi^0)$&-&-&-\\\hline
{$\Phi^0_{s1}$}&0.58&-10.14&1.84\\          
               &0.60&-3.46&2.80\\\hline     
{$\Phi_8^0$}&0.50&-23.42&1.44\\             
            &0.53&-4.93&2.52\\              
          \hline
&\multicolumn{3}{c}{$\mathcal{B-\bar B}$} \\\hline State&$\Lambda$
(GeV)&$E$ (MeV)& $r_{\mathrm{rms}}$ (fm)\\\hline
$\Omega^{\pm}(\Omega^0)$&-&-&-\\\hline {$\Omega^0_{s1}$}
                 &0.70&-9.24&1.84\\         
                 &0.72&-1.78&2.86\\ \hline  
{$\Omega_8^0$}&1.10&-7.13&1.95\\            
              &1.15&-23.34&1.23\\           
                       \end{tabular}
\caption{The bound state solutions for $\Phi^{\pm}(\Phi^0)$,
$\Phi_{s1}^0$ and $\Phi_8^0$ and $\Omega^{\pm}(\Omega^0)$,
$\Omega_{s1}^0$ and $\Omega_8^0$.\label{DD}}
\end{table}
\end{ruledtabular}
\end{center}

\begin{center}
\begin{ruledtabular}
\begin{table}[htb]
\begin{tabular}{c||ccccccccc}
&\multicolumn{3}{c}{$\mathcal{D-\bar D}$} \\\hline State&$\Lambda$
(GeV)&$E$ (MeV)& $r_{\mathrm{rms}}$ (fm)\\\hline
$\Phi^{\pm}(\Phi^0)$&-&-&-\\\hline
{$\Phi^0_{s1}$}&0.72&-23.00&1.30\\          
               &0.74&-8.16&1.90\\\hline     
{$\Phi_8^0$}&0.60&-37.20&1.17\\             
            &0.62&-11.39&1.77\\              
          \hline
&\multicolumn{3}{c}{$\mathcal{B-\bar B}$} \\\hline State&$\Lambda$
(GeV)&$E$ (MeV)& $r_{\mathrm{rms}}$ (fm)\\\hline
$\Omega^{\pm}(\Omega^0)$&-&-&-\\\hline {$\Omega^0_{s1}$}
                 &0.82&-16.70&1.44\\         
                 &0.83&-7.69&1.92\\ \hline  
{$\Omega_8^0$}&0.91&-11.84&1.66\\            
              &0.92&-21.27&1.32\\           
                       \end{tabular}
\caption{The bound state solutions for $\Phi^{\pm}(\Phi^0)$,
$\Phi_{s1}^0$ and $\Phi_8^0$ and $\Omega^{\pm}(\Omega^0)$,
$\Omega_{s1}^0$ and $\Omega_8^0$ if we increase the coupling
constants by a factor of two.\label{DD-2-times}}
\end{table}
\end{ruledtabular}
\end{center}

\begin{center}
\begin{ruledtabular}
\begin{table}[htb]
\begin{tabular}{c||ccccccccc}
&\multicolumn{3}{c}{$\mathcal{D-\bar D}$} \\\hline State&$\Lambda$
(GeV)&$E$ (MeV)& $r_{\mathrm{rms}}$ (fm)\\\hline
$\Phi^{\pm}(\Phi^0)$&-&-&-\\\hline
{$\Phi^0_{s1}$}&0.38&-24.49&1.50\\          
               &0.39&-18.89&1.63\\\hline     
{$\Phi_8^0$}&0.38&-15.38&1.78\\             
            &0.39&-10.61&2.02\\              
          \hline
&\multicolumn{3}{c}{$\mathcal{B-\bar B}$} \\\hline State&$\Lambda$
(GeV)&$E$ (MeV)& $r_{\mathrm{rms}}$ (fm)\\\hline
$\Omega^{\pm}(\Omega^0)$&-&-&-\\\hline {$\Omega^0_{s1}$}
                 &0.50&-30.69&1.29\\         
                 &0.52&-17.64&1.55\\ \hline  
{$\Omega_8^0$}&0.46&-29.24&1.37\\            
              &0.48&-14.15&1.73\\           
                       \end{tabular}
\caption{The bound state solutions for $\Phi^{\pm}(\Phi^0)$,
$\Phi_{s1}^0$ and $\Phi_8^0$ and $\Omega^{\pm}(\Omega^0)$,
$\Omega_{s1}^0$ and $\Omega_8^0$ if we reduce the coupling
constants by a factor of two.\label{DD-half-times}}
\end{table}
\end{ruledtabular}
\end{center}

\section{The $\mathcal{D}^\ast-\bar{\mathcal{D}}^\ast$ case}\label{secvv}

\subsection{The potential of the V-V system}

The possible quantum numbers of the S-wave V-V system are
$J^{P}=1^+, 1^+, 2^+$. The $C$ parity is $+$ for the neutral
states. The exchanged mesons include the pseudoscalar, vector and
$\sigma$ mesons. The exchange potential reads
\begin{eqnarray}
\mathcal{V}_\mathbb{P}^{[J]}(\textbf{q})\left[I_1^{\mathbb{P}\mathcal{D}^*_1\mathcal{D}^*_1}
,I_2^{\mathbb{P}\mathcal{ D}^*_2 \mathcal{
D}^*_2},m_{\mathcal{D}^*_1},m_{\mathcal{
D}^*_2},m_\mathbb{P}\right]&=&\frac{1}{4}\,g_{\mathcal{D^*D^*}\mathbb{P}}^2
\,\,I_1^{\mathbb{P}\mathcal{{D}}_1^*\mathcal{{D}}_1^*}\,I_2^{\mathbb{P}\mathcal{{D}}_2^*\mathcal{{D}}_2^*}
\,\mathcal{A}[J]\,\frac{\textbf{q}^2}{\textbf{q}^2+m_\mathbb{P}^2}\label{vvp1}
\end{eqnarray}
for the pseudoscalar meson exchange and
\begin{eqnarray}
&&\mathcal{V}^{[J]}_\mathbb{V}(\textbf{q})\left[I_1^{\mathbb{V}\mathcal{D}^*_1\mathcal{D}^*_1}
,I_2^{\mathbb{V}\mathcal{ D}^*_2 \mathcal{
D}^*_2},m_{\mathcal{D}^*_1},m_{\mathcal{
D}^*_2},m_\mathbb{V}\right]\nonumber\\&&\quad\quad=-I_1^{\mathbb{V}\mathcal{{D}}_1^*\mathcal{{D}}_1^*}
\,I_2^{\mathbb{V}\mathcal{{D}}_2^*\mathcal{{D}}_2^*}
\Bigg[\frac{g_{\mathcal{D^*D^*}\mathbb{V}}^2}{4m_{\mathcal{D}_1^*}m_{\mathcal{
D}_2^*}}
\,\mathcal{C}[J]\,\frac{\textbf{q}^2}{m_\mathbb{V}^2}+g_{\mathcal{D^*D^*}
\mathbb{V}}^2\,\mathcal{C}[J]\,\frac{1}{\mathbf{q}^2+m_{\mathbb
V}^2}+\frac{4f_{\mathcal{D^*D^*}\mathbb{V}}^2}{m_{\mathcal{D}_1^*}m_{\mathcal{
D}_2^*}}\,\mathcal{B}[J]\,\frac{\textbf{q}^2}
{\textbf{q}^2+m_\mathbb{V}^2}\Bigg]\label{vvp-2}
\end{eqnarray}
for the vector meson exchange. For the $\sigma$ exchange,
\begin{eqnarray}
\mathcal{V}^{[J]}_\sigma(\textbf{q}) &=&- {g_\sigma^2}
\,\mathcal{C}[J]\,\frac{1}{\textbf{q}^2+m_\sigma^2}\label{vvp-3}
\end{eqnarray}
with
\begin{eqnarray}
\mathcal{A}[J]&\equiv&\sum_{\lambda_1\lambda_2\lambda_3\lambda_4}\langle
1\lambda_1;1\lambda_2|J,m\rangle\langle1\lambda_3;1\lambda_4|J,m\rangle\frac{1}{\textbf{q}^2}
\left[\epsilon_1^{\lambda_{1}}\cdot
(\textbf{q}\times\epsilon_3^{*\lambda_{3}})\,\epsilon_2^{\lambda_{2}}\cdot(\textbf{q}\times\epsilon_4^{*\lambda_{4}})\right],\\
\mathcal{B}[J]&\equiv&\sum_{\lambda_1\lambda_2\lambda_3\lambda_4}\langle
1\lambda_1;1\lambda_2|J,m\rangle\langle1\lambda_3;1\lambda_4|J,m\rangle\frac{1}{\textbf{q}^2}[(\epsilon_1\cdot\textbf{q})
(\epsilon_2\cdot\textbf{q})(\epsilon^*_3\cdot\epsilon^*_4)+(c.t.s)],\\
\mathcal{C}[J]&\equiv&\sum_{\lambda_1\lambda_2\lambda_3\lambda_4}\langle
1\lambda_1;1\lambda_2|J,m\rangle\langle1\lambda_3;1
\lambda_4|J,m\rangle\left(\epsilon_1^{\lambda_1}
\cdot\epsilon_3^{*\lambda_3}\right)\left(\epsilon_2^{\lambda_2}\cdot\epsilon_4^{*\lambda_4}\right),
\end{eqnarray}
where $\epsilon_i$ is the polarization vector, and $c.t.s$ denotes
the cross-terms, i.e. $-(\epsilon_1\cdot\textbf{q})
(\epsilon_4^*\cdot\textbf{q})(\epsilon_2\cdot\epsilon_3^*)$,
$(\epsilon_3^*\cdot\textbf{q})
(\epsilon_4^*\cdot\textbf{q})(\epsilon_1\cdot\epsilon_2)$ and
$-(\epsilon_3^*\cdot\textbf{q})
(\epsilon_2\cdot\textbf{q})(\epsilon_1\cdot\epsilon_4^*)$. When
considering the different systems with $J^{P}=0^+, 1^+, 2^+$, we
impose the constraint on the scattering amplitudes that the
initial states and final states should have the same angular
momentum. Then we average the potential in the momentum space,
i.e., making the substitutions ${q_{x,y,z}^2\to {q^2}/{3}}$. The
coefficients of $\mathcal{A}(J)$, $\mathcal{B}(J)$ and
$\mathcal{C}(J)$ are listed into Table \ref{coeff}.
\begin{center}
\begin{ruledtabular}
\begin{table}[htb]
\begin{tabular}{c||ccc}
$J$&$\mathcal{A}(J)$&$\mathcal{B}(J)$&$\mathcal{C}(J)$\\\hline
0&2/3&4/3&1\\
1&1/3&2/3&1\\
2&-1/3&-2/3&1\\
\end{tabular}
\caption{The values of $\mathcal{A}(J)$, $\mathcal{B}(J)$ and
$\mathcal{C}(J)$ for the cases of $J^{P}=0^+, 1^+, 2^+$
systems.\label{coeff}}
\end{table}
\end{ruledtabular}
\end{center}
After Fourier transformation, we get the potentials in the
coordinate space
\begin{eqnarray}
\mathcal{V}_\mathbb{P}^{[J]}(r)\left[I_1^{\mathbb{P}\mathcal{D}^*_1\mathcal{D}^*_1}
,I_2^{\mathbb{P}\mathcal{D}^*_2 \mathcal{
D}^*_2},m_{\mathcal{D}^*_1},m_{\mathcal{
D}^*_2},m_\mathbb{P}\right]&=&\frac{1}{4}g_{\mathcal{D^*D^*}\mathbb{P}}^2
\,\,I_1^{\mathbb{P}\mathcal{{D}}_1^*\mathcal{{D}}_1^*}\,I_2^{\mathbb{P}\mathcal{{D}}_2^*\mathcal{{D}}_2^*}
\,\mathcal{A}[J]\,Z[\Lambda,m_\mathbb{P},r],
\end{eqnarray}
\begin{eqnarray}
&&\mathcal{V}^{[J]}_\mathbb{V}(r)\left[I_1^{\mathbb{V}\mathcal{D}^*_1\mathcal{D}^*_1}
,I_2^{\mathbb{V}\mathcal{ D}^*_2 \mathcal{
D}^*_2},m_{\mathcal{D}^*_1},m_{\mathcal{
D}^*_2},m_\mathbb{V}\right]\nonumber\\&&\quad\quad=-I_1^{\mathbb{V}\mathcal{{D}}_1^*\mathcal{{D}}_1^*}
\,I_2^{\mathbb{V}\mathcal{{D}}_2^*\mathcal{{D}}_2^*}
\Bigg[\frac{g_{\mathcal{D^*D^*}\mathbb{V}}^2}{4m_{\mathcal{D}_1^*}m_{\mathcal{
D}_2^*}}
\,\mathcal{C}[J]\,\frac{1}{m_\mathbb{V}^2}X[\Lambda,m_\mathbb{V},r]+g_{\mathcal{D^*D^*}
\mathbb{V}}^2\,\mathcal{C}[J]\,Y[\Lambda,m_\mathbb{V},r]
\nonumber\\&&\quad\quad\quad+\frac{4f_{\mathcal{D^*D^*}\mathbb{V}}^2}{m_{\mathcal{D}_1^*}m_{\mathcal{
D}_2^*}}\,\mathcal{B}[J]\,Z[\Lambda,m_\mathbb{V},r]\Bigg],
\end{eqnarray}
\begin{eqnarray}
\mathcal{V}_\sigma(r)^{[J]} &=&- {g_\sigma^2}
\,\mathcal{C}[J]\,Y[\Lambda,m_\sigma,r].
\end{eqnarray}
The total effective potentials for the $\Phi_s^{**\pm}$ and
$\Phi_s^0(\bar\Phi_s^0)$ states are
\begin{eqnarray}
\mathcal{V}(r)_{Total}^{\Phi_s^\pm
[J]}&=&\mathcal{V}^{[J]}_{\mathbb{P}}(r)
\left[\frac{1}{\sqrt{6}},-\frac{2}{\sqrt{6}},m_{D^{*0}},m_{D_s^*},m_{\eta}\right],
\end{eqnarray}
\begin{eqnarray}
\mathcal{V}(r)_{Total}^{\Phi_s^0(\bar\Phi_s^0)[J]}&=&\mathcal{V}^{[J]}_{\mathbb{P}}(r)
\left[\frac{1}{\sqrt{6}},-\frac{2}{\sqrt{6}},m_{D^{*+}},m_{D_s^*},m_{\eta}\right],
\end{eqnarray}
where only the $\eta$ meson exchange is allowed. Considering the
SU(2) symmetry, one further gets
\begin{eqnarray}
\mathcal{V}(r)_{Total}^{\Phi_s^\pm
[J]}&\approx&\mathcal{V}(r)_{Total}^{\Phi_s^0(\bar\Phi_s^0)[J]}=
-\frac{1}{12}\,g_{\mathcal{D^*D^*}\mathbb{P}}^2
\,\,\mathcal{A}[J]\,Z[\Lambda,m_\eta,r].\label{vv-1}
\end{eqnarray}
For the $\Phi_{s1}^{**0}$ state, the potential is
\begin{eqnarray}
\mathcal{V}(r)_{Total}^{\Phi_{s1}^{**0}[J]}&=&\mathcal{V}^{[J]}_{\mathbb{P}}(r)
\left[-\frac{2}{\sqrt{6}},-\frac{2}{\sqrt{6}},m_{D_s^{*}},m_{D_s^*},m_{\eta}\right]+
\mathcal{V}^{[J]}_{\mathbb{V}}(r)
\left[1,1,m_{D^{*}_s},m_{D^{*}_s},m_{\phi}\right]\nonumber\\
&=&\frac{1}{6}g_{\mathcal{D^*D^*}\mathbb{P}}^2\,\mathcal{A}[J]Z[\Lambda,m_{\eta},r]
-
\Bigg[\frac{g_{\mathcal{D^*D^*}\mathbb{V}}^2}{4m_{\mathcal{D}_s^*}^2m_\phi^2}
\,\mathcal{C}[J]\,X[\Lambda,m_\phi,r]+g_{\mathcal{D^*D^*}
\mathbb{V}}^2\,\mathcal{C}[J]\,Y[\Lambda,m_\phi,r]
\nonumber\\&&+\frac{4f_{\mathcal{D^*D^*}\mathbb{V}}^2}{m_{\mathcal{D}_s^*}^2}\,\mathcal{B}[J]\,
Z[\Lambda,m_\phi,r]\Bigg].\label{vv-4}
\end{eqnarray}
For $\Phi^{**\pm}$, the exchange potential reads
\begin{eqnarray}
\mathcal{V}(r)_{Total}^{\Phi^{**\pm}[J]}&=&\mathcal{V}^{[J]}_{\mathbb{P}}(r)
\left[\frac{1}{\sqrt{2}},-\frac{1}{\sqrt{2}},m_{D^{*0}},m_{D^{*+}},m_{\pi}\right]+\mathcal{V}^{[J]}_{\mathbb{P}}(r)
\left[\frac{1}{\sqrt{6}},\frac{1}{\sqrt{6}},m_{D^{*0}},m_{D^{*+}},m_{\eta}\right]\nonumber\\&&+
\mathcal{V}^{[J]}_{\mathbb{V}}(r)
\left[\frac{1}{\sqrt{2}},-\frac{1}{\sqrt{2}},m_{D^{*0}},m_{D^{*+}},m_{\rho}\right]+\mathcal{V}^{[J]}_{\mathbb{V}}(r)
\left[\frac{1}{\sqrt{2}},\frac{1}{\sqrt{2}},m_{D^{*0}},m_{D^{*+}},m_{\omega}\right]\nonumber\\&&+
\mathcal{V}_\sigma(r)^{[J]}\nonumber\\&\approx&
-g_{\mathcal{D^*D^*}\mathbb{P}}^2\,\mathcal{A}[J]\,\left[\frac{Z[\Lambda,m_{\pi},r]}{8}
-\frac{Z[\Lambda,m_{\eta},r]}{24}\right]-{g_\sigma^2}
\,\mathcal{C}[J]\,Y[\Lambda,m_\sigma,r]\nonumber\\&&+ \frac{1}{2}
\Bigg[\frac{g_{\mathcal{D^*D^*}\mathbb{V}}^2}{4m_{\mathcal{D}^*}^2m_\rho^2}
\,\mathcal{C}[J]\,X[\Lambda,m_\rho,r]+g_{\mathcal{D^*D^*}
\mathbb{V}}^2\,\mathcal{C}[J]\,Y[\Lambda,m_\rho,r]
+\frac{4f_{\mathcal{D^*D^*}\mathbb{V}}^2}{m_{\mathcal{D}^*}^2}\,\mathcal{B}[J]\,Z[\Lambda,m_\rho,r]\Bigg]\nonumber\\&&
- \frac{1}{2}
\Bigg[\frac{g_{\mathcal{D^*D^*}\mathbb{V}}^2}{4m_{\mathcal{D}^*}^2m_\omega^2}
\,\mathcal{C}[J]\,X[\Lambda,m_\omega,r]+g_{\mathcal{D^*D^*}
\mathbb{V}}^2\,\mathcal{C}[J]\,Y[\Lambda,m_\omega,r]
+\frac{4f_{\mathcal{D^*D^*}\mathbb{V}}^2}{m_{\mathcal{D}^*}^2}\,\mathcal{B}[J]\,Z[\Lambda,m_\omega,r]\Bigg].\nonumber\\&&
\label{vv-2}
\end{eqnarray}
The potential for the $\Phi^{**0}(\Phi_8^{**0})$ state is
\begin{eqnarray}
\mathcal{V}(r)^{\Phi^{**0}(\Phi_8^{**0})[J]}_{Total}&=&
\frac{1}{2}\Bigg\{\mathcal{V}_{\mathbb{P}}^{[J]}(r)
\left[\frac{1}{\sqrt{2}},\frac{1}{\sqrt{2}},m_{D^{*0}},m_{D^{*0}},m_{\pi^0}\right]
\mp2\mathcal{V}_{\mathbb{P}}^{[J]}(r)
\left[1,1,{m_{D^{*0}}},m_{D^{*+}},m_{\pi^\pm}\right]
\nonumber\\&&+\mathcal{V}_{\mathbb{P}}^{[J]}(r)
\left[-\frac{1}{\sqrt{2}},-\frac{1}{\sqrt{2}},m_{D^{*+}},m_{D^{*-}},m_{\pi^0}\right]+
\mathcal{V}_{\mathbb{P}}^{[J]}(r)
\left[\frac{1}{\sqrt{6}},\frac{1}{\sqrt{6}},m_{D^{*0}},m_{D^{*0}},m_{\eta}\right]\nonumber\\&&
+ \mathcal{V}_{\mathbb{P}}^{[J]}(r)
\left[\frac{1}{\sqrt{6}},\frac{1}{\sqrt{6}},m_{D^{*+}},m_{D^{*-}},m_{\eta}\right]
\nonumber\\&&+ \mathcal{V}_{\mathbb{V}}^{[J]}(r)
\left[\frac{1}{\sqrt{2}},\frac{1}{\sqrt{2}},m_{D^{*0}},m_{D^{*0}},m_{\rho^0}\right]
\mp2\mathcal{V}_{\mathbb{V}}^{[J]}(r)
\left[1,1,{m_{D^{*0}}},m_{D^{*+}},m_{\rho^\pm}\right]
\nonumber\\&&+\mathcal{V}_{\mathbb{V}}^{[J]}(r)
\left[-\frac{1}{\sqrt{2}},-\frac{1}{\sqrt{2}},m_{D^{*+}},m_{D^{*-}},m_{\rho^0}\right]+
\mathcal{V}_{\mathbb{V}}^{[J]}(r)
\left[\frac{1}{\sqrt{2}},\frac{1}{\sqrt{2}},m_{D^{*0}},m_{D^{*0}},m_{\omega}\right]\nonumber\\&&
+ \mathcal{V}_{\mathbb{V}}^{[J]}(r)
\left[\frac{1}{\sqrt{2}},\frac{1}{\sqrt{2}},m_{D^{*+}},m_{D^{*-}},m_{\omega}\right]
 + 2\mathcal{V}_{\sigma}^{[J]}(r)
 \Bigg\}\nonumber\\&\approx&
g_{\mathcal{D^*D^*}\mathbb{P}}^2\,\mathcal{A}[J]\,\left[\frac{(1\mp2)}{8}Z[\Lambda,m_{\pi},r]
+\frac{Z[\Lambda,m_{\eta},r]}{24}\right]-{g_\sigma^2}
\,\mathcal{C}[J]\,Y[\Lambda,m_\sigma,r]\nonumber\\&&-
\frac{(1\mp2)}{2}
\Bigg[\frac{g_{\mathcal{D^*D^*}\mathbb{V}}^2}{4m_{\mathcal{D}^*}^2m_\rho^2}
\,\mathcal{C}[J]\,X[\Lambda,m_\rho,r]+g_{\mathcal{D^*D^*}
\mathbb{V}}^2\,\mathcal{C}[J]\,Y[\Lambda,m_\rho,r]
+\frac{4f_{\mathcal{D^*D^*}\mathbb{V}}^2}{m_{\mathcal{D}^*}^2}\,\mathcal{B}[J]\,Z[\Lambda,m_\rho,r]\Bigg]\nonumber\\&&
- \frac{1}{2}
\Bigg[\frac{g_{\mathcal{D^*D^*}\mathbb{V}}^2}{4m_{\mathcal{D}^*}^2m_\omega^2}
\,\mathcal{C}[J]\,X[\Lambda,m_\omega,r]+g_{\mathcal{D^*D^*}
\mathbb{V}}^2\,\mathcal{C}[J]\,Y[\Lambda,m_\omega,r]
+\frac{4f_{\mathcal{D^*D^*}\mathbb{V}}^2}{m_{\mathcal{D}^*}^2}\,\mathcal{B}[J]\,Z[\Lambda,m_\omega,r]\Bigg],\nonumber\\\label{vv-3}
\end{eqnarray}
where $\mp$ corresponds to $\Phi^{**0}(\Phi_8^{**0})$
respectively. As a cross-check, the potential for $\Phi^{**0}$ is
the same as that for $\Phi^{**\pm}$ from the SU(2) symmetry. For
the V-V systems, we need only consider four independent types of
potentials corresponding to
$[\Phi_{s}^{**\pm},\Phi_s^{**0},\bar\Phi_{s}^{**0}]$,
$[\Phi^{**\pm},\Phi^{**0}]$, $\Phi_{8}^{**0}$ and
$\Phi_{s1}^{**0}$ listed in Eqs. (\ref{vv-1}), (\ref{vv-2}),
(\ref{vv-3}), (\ref{vv-4}) respectively when we consider the SU(2)
symmetry. For states with different quantum numbers, we simply
change the values of $\mathcal{A}[J]$, $\mathcal{B}[J]$ and
$\mathcal{C}[J]$ in Table \ref{coeff}.

\subsection{Numerical results for the V-V system}\label{result-vv}

\begin{figure}[htb]\begin{center}
\begin{tabular}{ccc}
\scalebox{0.8}{\includegraphics{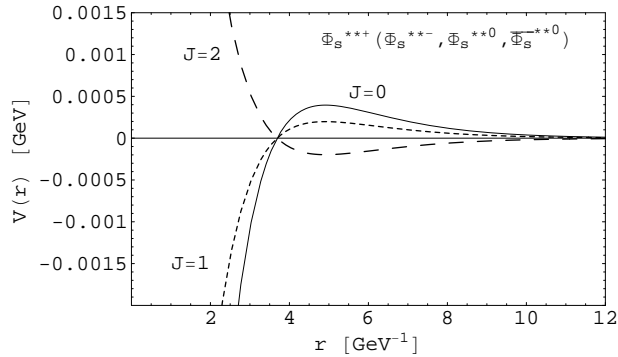}}\\
\end{tabular}
\caption{The effective potential of
$\Phi_s^{**\pm}(\Phi_{s}^{**0},\bar{\Phi}_{s}^{**0})$.
\label{VV-potential-1}}\end{center}
\end{figure}

\begin{center}
\begin{figure}[htb]
\begin{tabular}{ccc}
\scalebox{0.55}{\includegraphics{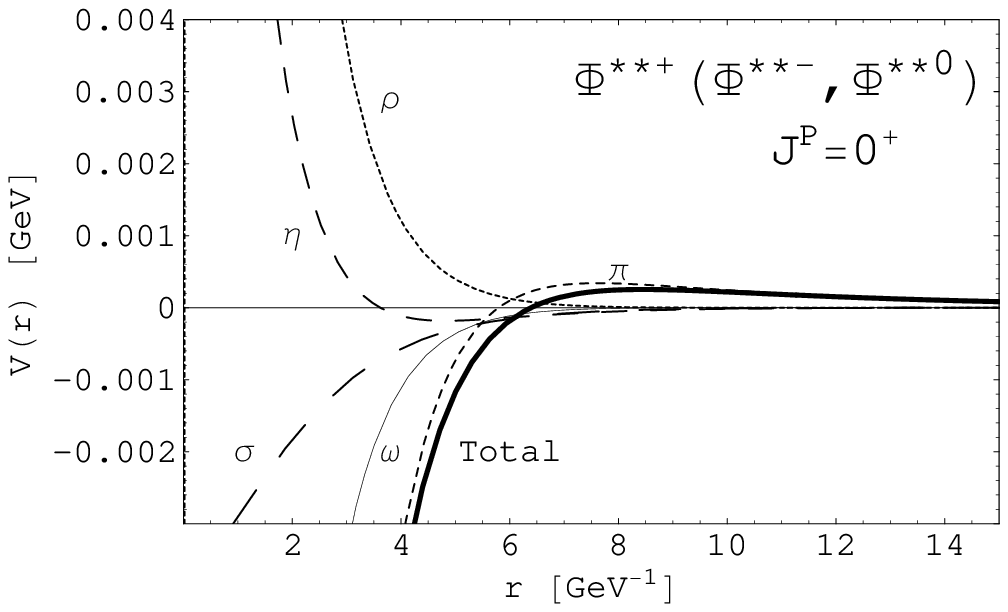}}&
\scalebox{0.55}{\includegraphics{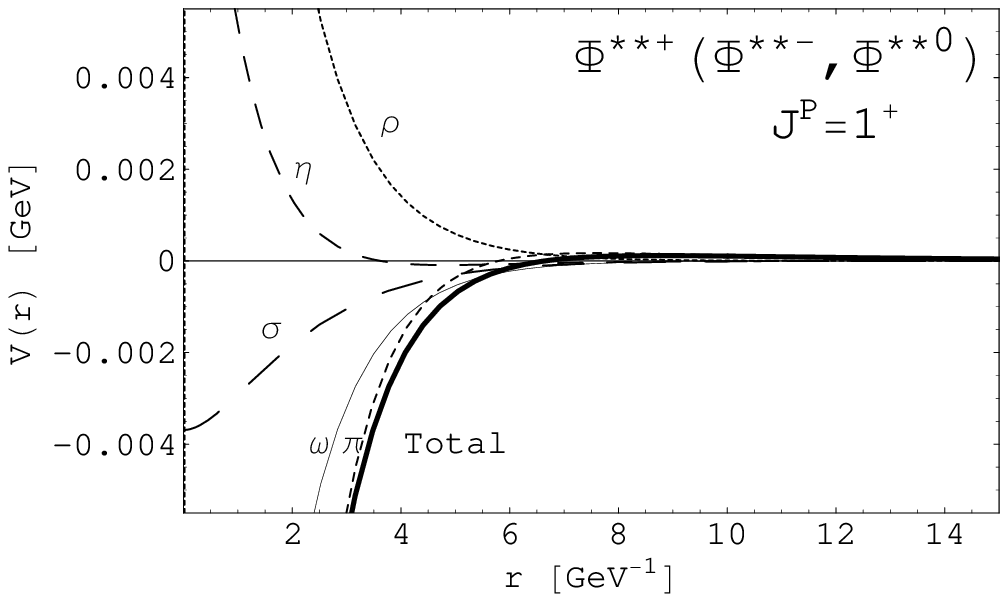}}&
\scalebox{0.55}{\includegraphics{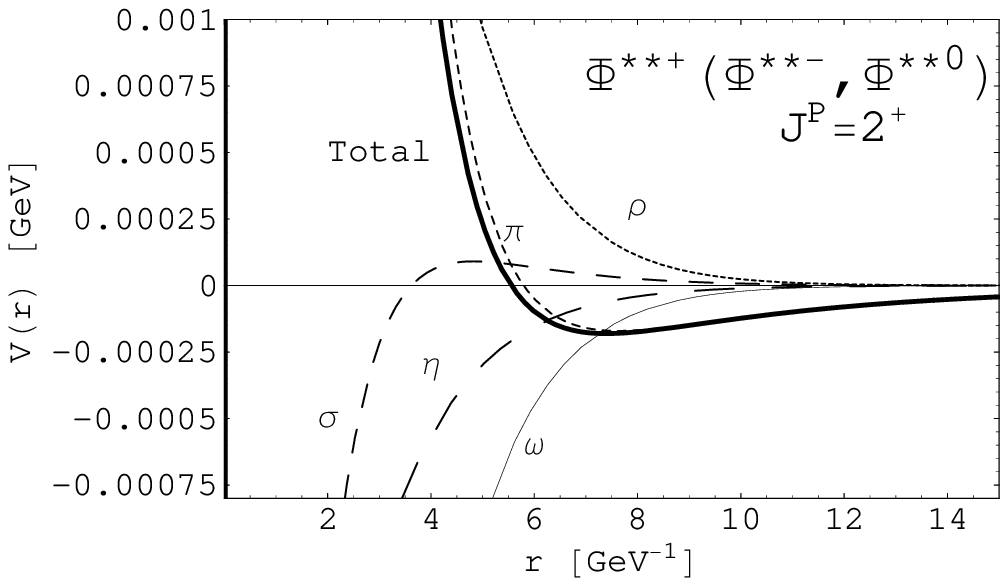}}\\(a)&(b)&(c)\\
\end{tabular}
\caption{The effective potentials of $\Phi^{**\pm}(\Phi^{**0})$
with different quantum numbers. The thick solid line is the total
effective potential. \label{VV-potential-2}}
\end{figure}\end{center}

\begin{figure}[htb]\begin{center}
\begin{tabular}{ccc}
\scalebox{0.55}{\includegraphics{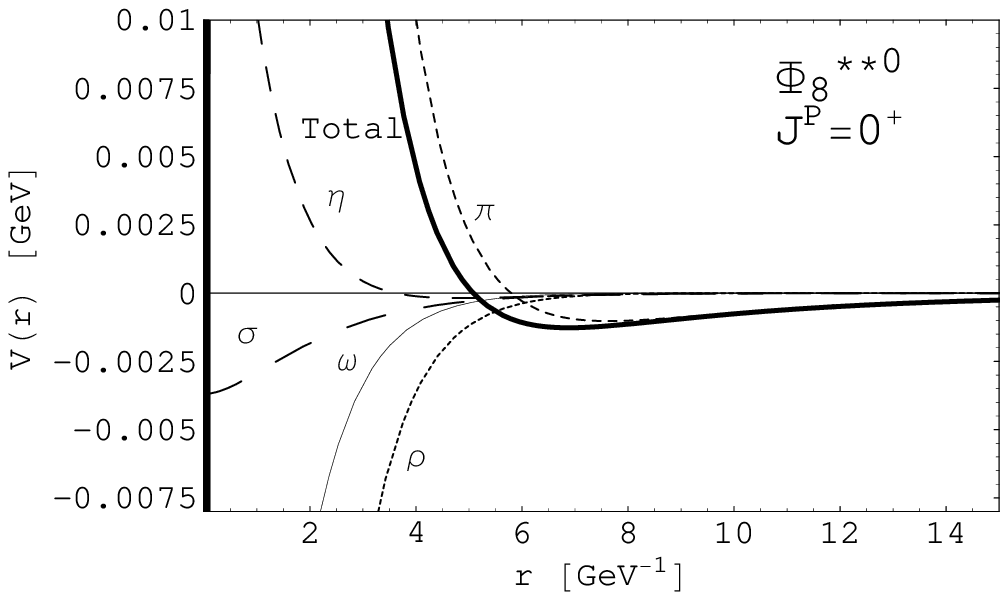}}&
\scalebox{0.55}{\includegraphics{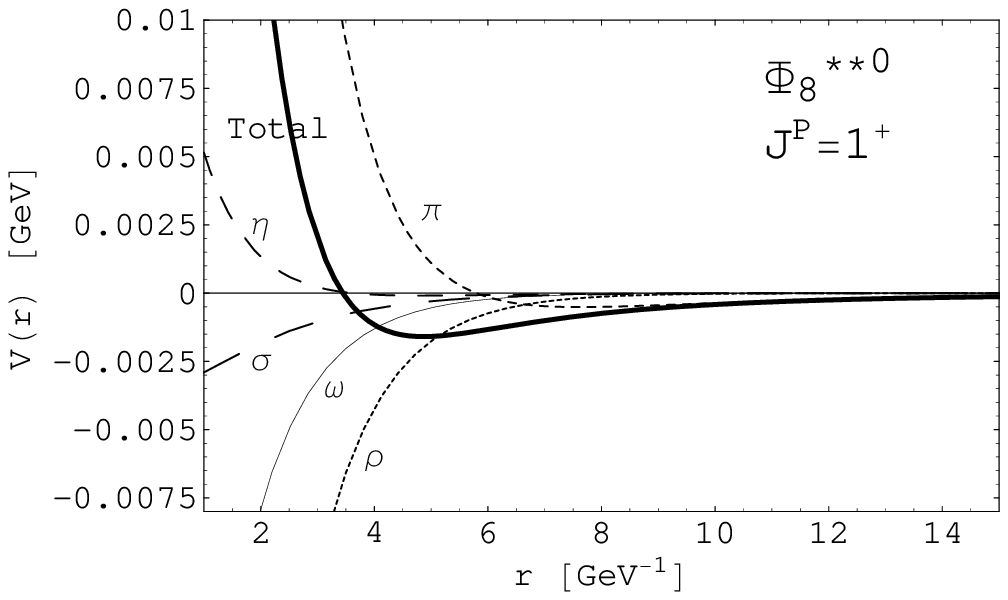}}&
\scalebox{0.55}{\includegraphics{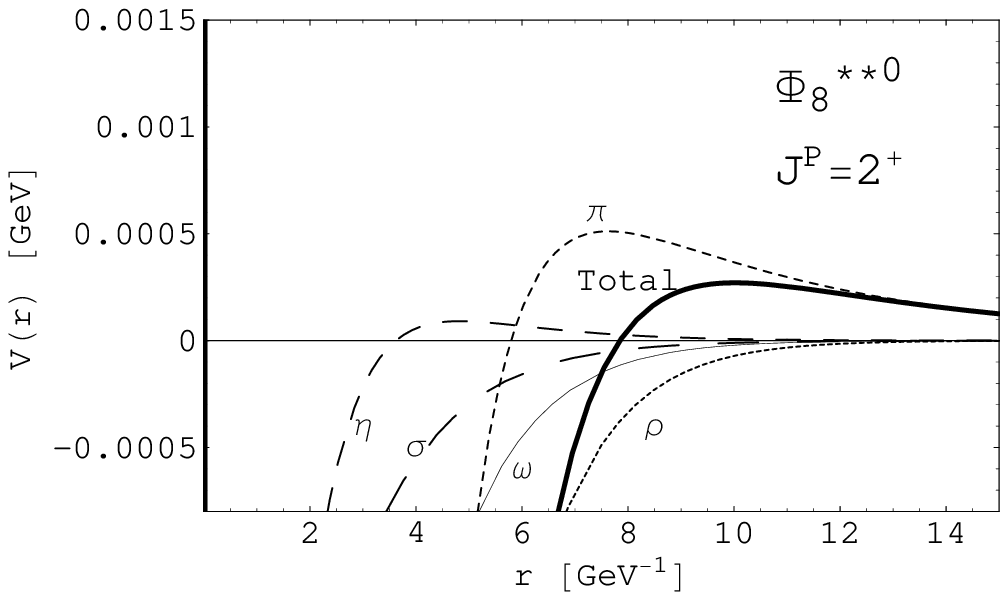}}\\(a)&(b)&(c)\\
\end{tabular}
\caption{The shape of the exchange potential of $\Phi_8^{**0}$.
The thick solid line is the total effective potential.
\label{VV-potential-3}}\end{center}
\end{figure}

\begin{figure}[htb]\begin{center}
\begin{tabular}{ccc}
\scalebox{0.55}{\includegraphics{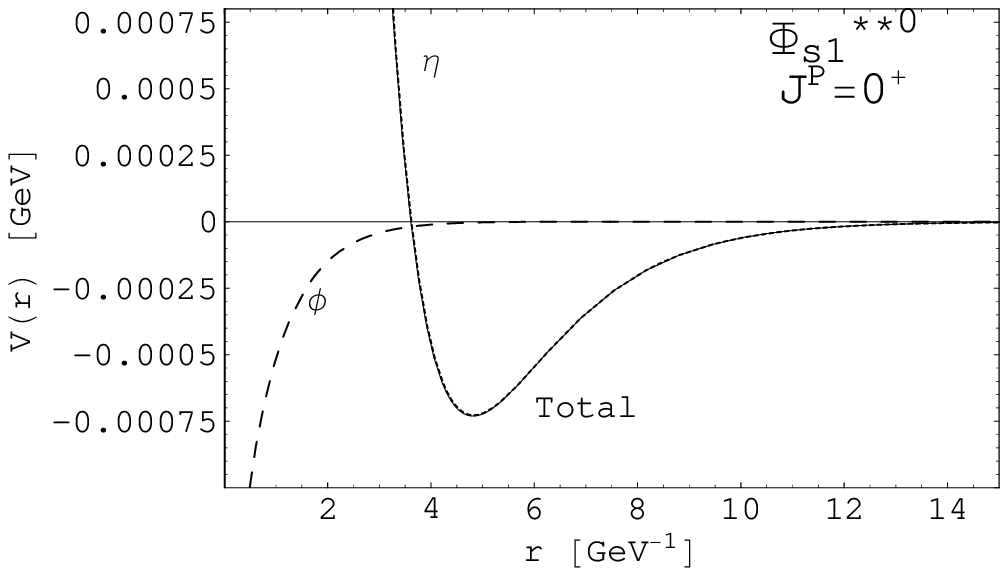}}&
\scalebox{0.55}{\includegraphics{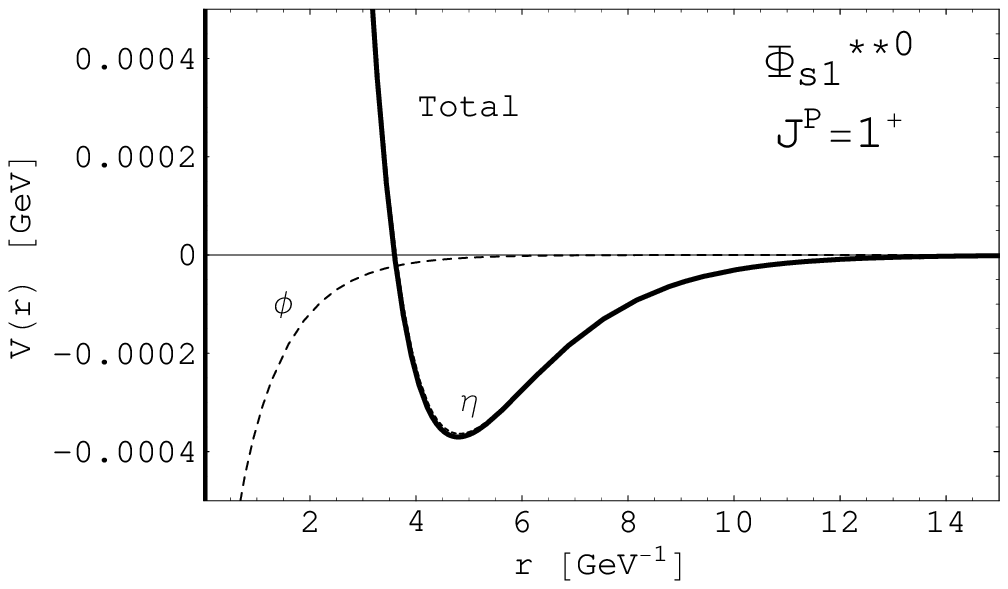}}&
\scalebox{0.55}{\includegraphics{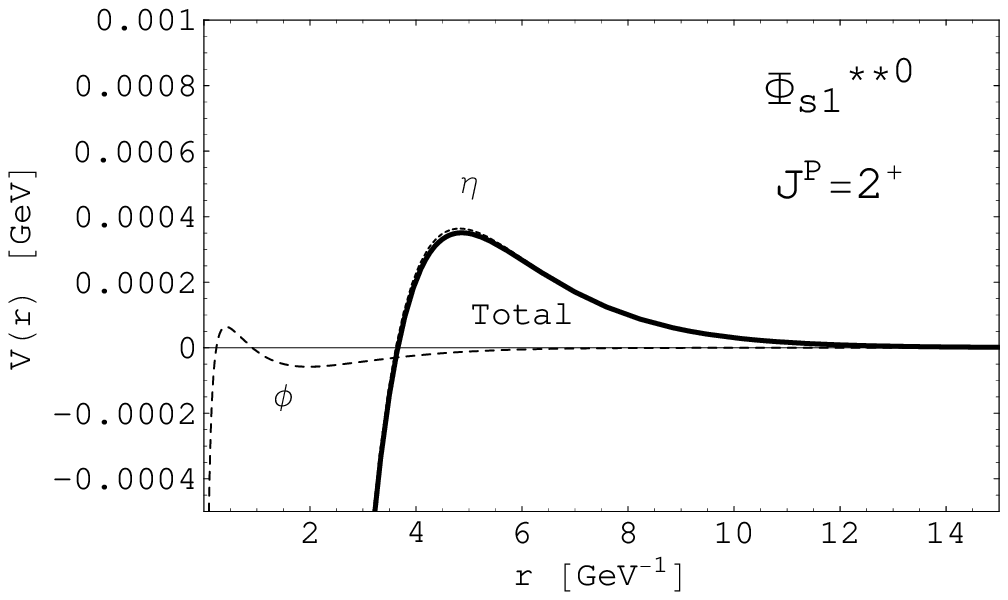}}\\(a)&(b)&(c)\\
\end{tabular}
\caption{The variation of the effective potential of
$\Phi_{s1}^{**0}$ with $r$. The thick solid line is the total
effective potential. \label{VV-potential-4}}\end{center}
\end{figure}

The quantum numbers of the S-wave V-V system are $J^P=0^+, 1^+,
2^+$. The variation of the oscillating effective potential of
$\Phi_{s}^{**\pm}(\Phi_{s}^{**0},\bar{\Phi}_{s}^{**0})$ with $r$
is shown in Fig. \ref{VV-potential-1}, where only the $\eta$ meson
exchange is allowed. The effective potentials of
$\Phi^{*\pm}(\Phi^{*0})$ and $\Phi_8^{**0}$ with $J^P=0^+, 1^+,
2^+$ are presented in Fig.
\ref{VV-potential-2}-\ref{VV-potential-3}, where the exchanged
mesons include $\pi$, $\eta$, $\rho$, $\omega$ and $\sigma$. For
$\Phi_{s1}^{**0}$, its effective potential is shown in Fig.
\ref{VV-potential-4}, where both the $\eta$ and $\phi$ meson
exchange is allowed.


With the above effective potentials, we solve the Schr\"{o}dinger
equation to find the bound state solutions for the V-V system.
Numerical results are collected in Table \ref{DsDs}. For the
$\mathcal{D^*-D^*}$ states with $J^{P}=0^+, 1^+$, a bound state
exists only for $\Lambda$ much larger than $1$ GeV. Especially for
$\Phi_s^{**\pm}(\Phi_s^{**0},\bar\Phi_{s}^{**0})$ and
$\Phi^{**\pm}(\Phi^{**0})$ with $J^P=1^+$, the existence of a
bound state requires the values of $\Lambda$ be larger than 8 GeV.
For $\Phi_s^{**\pm}(\Phi_s^{**0},\bar\Phi_{s}^{**0})$ and
$\Phi^{**\pm}(\Phi^{**0})$ with $J^P=2^+$, no bound states exists
for $\Lambda< 10$ GeV. The cutoff parameter $\Lambda$ is a typical
hadronic scale, which is generally expected to be around $1\sim 2$
GeV. If $\Lambda$ is much larger than 2 GeV or much smaller than 1
GeV in order to form a bound state, we tend to conclude that there
does not exist a heavy molecular state for these systems. We find
bound state solutions for $\Phi_{8}^{**0}$ and $\Phi_{s1}^{**0}$
with $J^P=0^+,1^+,2^+$.

\begin{center}
\begin{figure}[htb]
\begin{tabular}{c}
\scalebox{0.6}{ \includegraphics{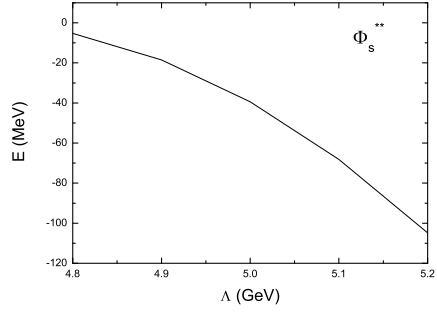}}
\end{tabular} \caption{The dependence of binding energy $E$ on $\Lambda$ for $\Phi_{s}^{**}$. \label{VV-Phi_s}}
\end{figure}
\end{center}

In Table \ref{DsDs}, we also give the numerical results for the
$\mathcal{B^*-\bar B^*}$ system. There do not exist molecular
states $\Omega_s^{**\pm}(\Omega_s^{**0},\bar\Omega_{s}^{**0})$ and
$\Omega^{**\pm}(\Omega^{**0})$ with $J^P=2^+$ since we can not
find bound state solutions for $\Lambda=0\sim 10$ GeV. Different
from the $\mathcal{D^*-\bar D^*}$ system, there exist possible
molecular states
$\Omega_{s}^{**\pm}(\Omega_{s}^{**0},\bar{\Omega}_{s}^{**0})$ with
$J^{P}=0^+, 1^+$ with more reasonable $\Lambda$ close to $1$ GeV.
We also find the bound state solutions for $\Omega_{8}^{**0}$ and
$\Omega_{s1}^{**0}$. The reason is simple. The larger the reduced
mass for $\mathcal{B^*-\bar B^*}$, the lower the kinetic energy,
the easier to form a molecular state. With $\Phi_{s}^{**}$ as an
example, we illustrate the dependence of $E$ on $\Lambda$ in Fig.
\ref{VV-Phi_s}. When all coupling constants are enlarged by a
factor of two, the numerical results are collected in Table
\ref{DsDs-2times}. When all coupling constants are reduced by a
factor of two, the results are collected in Table \ref{DsDs-half}.

\begin{center}
\begin{ruledtabular}
\begin{table}[htb]
\begin{tabular}{c||ccc|ccc|ccccc}
&\multicolumn{8}{c}{$\mathcal{D^*-\bar D^*}$} \\\hline

&\multicolumn{3}{c}{$J^P=0^+$}&\multicolumn{3}{c}{$J^P=1^+$}&\multicolumn{3}{c}{$J^P=2^+$}\\\hline

 State&$\Lambda$
(GeV)&$E$ (MeV)& $r_{\mathrm{rms}}$ (fm)&$\Lambda$ (GeV)&$E$
(MeV)& $r_{\mathrm{rms}}$ (fm)&$\Lambda$ (GeV)&$E$ (MeV)&
$r_{\mathrm{\mathrm{rms}}}$ (fm)\\\hline

$\Phi_s^{**\pm}(\Phi_s^{**0},\bar\Phi_{s}^{**0})$&4.80&-5.26&1.92&8.70&-1.60&2.56&-&-&-\\
&5.00&-39.42&0.72&8.90&-21.21&0.96&-&-&-\\\hline

$\Phi^{**\pm}(\Phi^{**0})$&4.00&-5.21&1.99&9.80&-8.40&1.61&-&-&-\\
&4.20&-19.02&1.08&10.00&-16.53&1.19&-&-&-\\\hline

$\Phi_8^{**0}$&0.52&-18.32&1.48&0.50&-37.61&1.18&0.52&-11.71&1.85\\
              &0.53&-8.61&1.95&0.52&-15.03&1.63&0.50&-24.01&1.47\\\hline

$\Phi_{s1}^{**0}$ &0.63 &-12.91&1.55 &0.60&-17.27&1.45&0.54&-23.64&1.46\\
                  &0.64&-7.13&1.93&0.62&-6.56&2.11&0.56&-13.01&1.77\\\hline

&\multicolumn{8}{c}{$\mathcal{B^*-\bar B^*}$} \\\hline

&\multicolumn{3}{c}{$J^P=0^+$}&\multicolumn{3}{c}{$J^P=1^+$}&\multicolumn{3}{c}{$J^P=2^+$}\\\hline

 State&$\Lambda$
(GeV)&$E$ (MeV)& $r_{\mathrm{rms}}$ (fm)&$\Lambda$ (GeV)&$E$
(MeV)& $r_{\mathrm{rms}}$ (fm)&$\Lambda$ (GeV)&$E$ (MeV)&
$r_{\mathrm{rms}}$ (fm)\\\hline

$\Omega_s^{**\pm}(\Omega_s^{**0},\bar\Omega_{s}^{**0})$&2.30&-0.57&4.76&3.80&-3.98&2.21&-&-&-\\
&2.40&-14.90&1.17&3.90&-17.91&1.06&-&-&-\\\hline

$\Omega^{**\pm}(\Omega^{**0})$&1.50&-10.17&1.53&2.70&-3.60&2.41&-&-&-\\
&1.60&-21.75&1.10&2.90&-15.55&1.22&-&-&-\\\hline

$\Omega_8^{**0}$&0.57&-23.16&1.36&0.57&-32.70&1.23&0.61&-18.07&1.49\\
              &0.58&-6.42&2.28&0.59&-5.65&2.39&0.63&-8.39&1.93\\\hline

$\Omega_{s1}^{**0}$ & 0.74&-12.12&1.56 &0.70&-33.5&1.09&0.64&-35.37&1.25\\
                  &0.75&-4.48&2.37&0.72&-12.68&1.58&0.65&-27.12&1.35\\

            \end{tabular}
\caption{The numerical results for the $\mathcal{D^*-\bar D^*}$
and $\mathcal{B^*-\bar B^*}$ systems.\label{DsDs}}
\end{table}
\end{ruledtabular}
\end{center}

\begin{center}
\begin{ruledtabular}
\begin{table}[htb]
\begin{tabular}{c||ccc|ccc|ccccc}
&\multicolumn{8}{c}{$\mathcal{D^*-\bar D^*}$} \\\hline

&\multicolumn{3}{c}{$J^P=0^+$}&\multicolumn{3}{c}{$J^P=1^+$}&\multicolumn{3}{c}{$J^P=2^+$}\\\hline

 State&$\Lambda$
(GeV)&$E$ (MeV)& $r_{\mathrm{rms}}$ (fm)&$\Lambda$ (GeV)&$E$
(MeV)& $r_{\mathrm{rms}}$ (fm)&$\Lambda$ (GeV)&$E$ (MeV)&
$r_{\mathrm{\mathrm{rms}}}$ (fm)\\\hline

$\Phi_s^{**\pm}(\Phi_s^{**0},\bar\Phi_{s}^{**0})$&1.90&-16.07&1.14&2.90&-14.48&1.18&-&-&-\\
&1.85&-4.7&2.05&2.80&-1.14&3.74&-&-&-\\\hline

$\Phi^{**\pm}(\Phi^{**0})$&1.00&-5.42&2.06&1.80&-6.28&1.89&-&-&-\\
&1.20&-36.75&0.92&2.00&-24.25&1.04&-&-&-\\\hline

$\Phi_8^{**0}$&0.58&-45.86&1.05&0.60&-15.53&1.62&0.65&-32.94&1.15\\
              &0.59&-13.02&1.73&0.61&-3.64&2.64&0.66&-27.10&1.21\\\hline

$\Phi_{s1}^{**0}$ &0.76 &-39.20&0.97 &0.74&-38.98&1.01&0.70&-23.96&1.38\\
                  &0.78&-7.90&1.82&0.76&-12.73&1.55&0.72&-11.31&1.78\\\hline

&\multicolumn{8}{c}{$\mathcal{B^*-\bar B^*}$} \\\hline

&\multicolumn{3}{c}{$J^P=0^+$}&\multicolumn{3}{c}{$J^P=1^+$}&\multicolumn{3}{c}{$J^P=2^+$}\\\hline

 State&$\Lambda$
(GeV)&$E$ (MeV)& $r_{\mathrm{rms}}$ (fm)&$\Lambda$ (GeV)&$E$
(MeV)& $r_{\mathrm{rms}}$ (fm)&$\Lambda$ (GeV)&$E$ (MeV)&
$r_{\mathrm{rms}}$ (fm)\\\hline

$\Omega_s^{**\pm}(\Omega_s^{**0},\bar\Omega_{s}^{**0})$&1.20&-9.26&1.50&1.60&-6.27&1.79&-&-&-\\
&1.30&-79.29&0.60&1.70&-44.38&0.73&-&-&-\\\hline

$\Omega^{**\pm}(\Omega^{**0})$&0.50&-17.96&1.41&0.80&-7.74&1.80&-&-&-\\
&0.48&-16.36&1.48&0.90&-25.86&1.11&-&-&-\\\hline

$\Omega_8^{**0}$&0.60&-81.06&0.89&0.62&-46.66&1.15&0.70&-245.16&0.50\\
              &0.61&-10.73&2.15&0.63&-11.49&1.98&0.71&-352.21&0.49\\\hline

$\Omega_{s1}^{**0}$ & 0.83&-25.32&1.17 &0.82&-34.93&1.06&0.80&-26.93&1.27\\
                  &0.84&-5.08&2.18&0.83&-15.01&1.47&0.82&-11.66&1.69\\

            \end{tabular}
\caption{The numerical results for the $\mathcal{D^*-\bar D^*}$
and $\mathcal{B^*-\bar B^*}$ systems after increasing the coupling
constants by a factor of two.\label{DsDs-2times}}
\end{table}
\end{ruledtabular}
\end{center}

\begin{center}
\begin{ruledtabular}
\begin{table}[htb]
\begin{tabular}{c||ccc|ccc|ccccc}
&\multicolumn{8}{c}{$\mathcal{D^*-\bar D^*}$} \\\hline

&\multicolumn{3}{c}{$J^P=0^+$}&\multicolumn{3}{c}{$J^P=1^+$}&\multicolumn{3}{c}{$J^P=2^+$}\\\hline

 State&$\Lambda$
(GeV)&$E$ (MeV)& $r_{\mathrm{rms}}$ (fm)&$\Lambda$ (GeV)&$E$
(MeV)& $r_{\mathrm{rms}}$ (fm)&$\Lambda$ (GeV)&$E$ (MeV)&
$r_{\mathrm{\mathrm{rms}}}$ (fm)\\\hline

$\Phi_s^{**\pm}(\Phi_s^{**0},\bar\Phi_{s}^{**0})$&-&-&-&-&-&-&-&-&-\\\hline

$\Phi^{**\pm}(\Phi^{**0})$&-&-&-&-&-&-&-&-&-\\\hline

$\Phi_8^{**0}$&0.40&-18.40&1.57&0.40&-13.09&1.82&0.39&-10.14&2.10\\
              &0.41&-12.23&1.81&0.41&-8.43&2.12&0.40&-6.67&2.42\\\hline

$\Phi_{s1}^{**0}$ &0.46 &-6.43&2.21 &0.40&-27.87&1.37&0.40&-12.05&1.95\\
                  &0.47&-3.70&2.76&0.42&-16.15&1.63&0.41&-8.61&2.18\\\hline

&\multicolumn{8}{c}{$\mathcal{B^*-\bar B^*}$} \\\hline

&\multicolumn{3}{c}{$J^P=0^+$}&\multicolumn{3}{c}{$J^P=1^+$}&\multicolumn{3}{c}{$J^P=2^+$}\\\hline

 State&$\Lambda$
(GeV)&$E$ (MeV)& $r_{\mathrm{rms}}$ (fm)&$\Lambda$ (GeV)&$E$
(MeV)& $r_{\mathrm{rms}}$ (fm)&$\Lambda$ (GeV)&$E$ (MeV)&
$r_{\mathrm{rms}}$ (fm)\\\hline

$\Omega_s^{**\pm}(\Omega_s^{**0},\bar\Omega_{s}^{**0})$&6.80&-9.33&1.43&-&-&-&-&-&-\\
&6.90&-23.17&0.93&-&-&-&-&-&-\\\hline

$\Omega^{**\pm}(\Omega^{**0})$&6.10&-7.37&1.64&-&-&-&-&-&-\\
&6.20&-14.64&1.20&-&-&-&-&-&-\\\hline

$\Omega_8^{**0}$&0.48&-26.88&1.31&0.48&-19.59&1.51&0.48&-11.34&1.92\\
              &0.50&-8.85&1.97&0.50&-6.57&2.25&0.50&-4.18&2.75\\\hline

$\Omega_{s1}^{**0}$ & 0.58&-11.18&1.68 &0.53&-29.28&1.23&0.50&-18.73&1.60\\
                  &0.60&-3.11&2.85&0.55&-15.49&1.54&0.52&-9.95&1.97\\

            \end{tabular}
\caption{The numerical results for the $\mathcal{D^*-\bar D^*}$
and $\mathcal{B^*-\bar B^*}$ systems after reducing the coupling
constants by a factor of two.\label{DsDs-half}}
\end{table}
\end{ruledtabular}
\end{center}

\section{The $\mathcal{D}^\ast-\bar{\mathcal{D}}$ case}\label{secpv}

\subsection{The potential of the P-V system}

\begin{center}
\begin{figure}[htb]
\begin{tabular}{ccc}
\scalebox{0.8}{\includegraphics{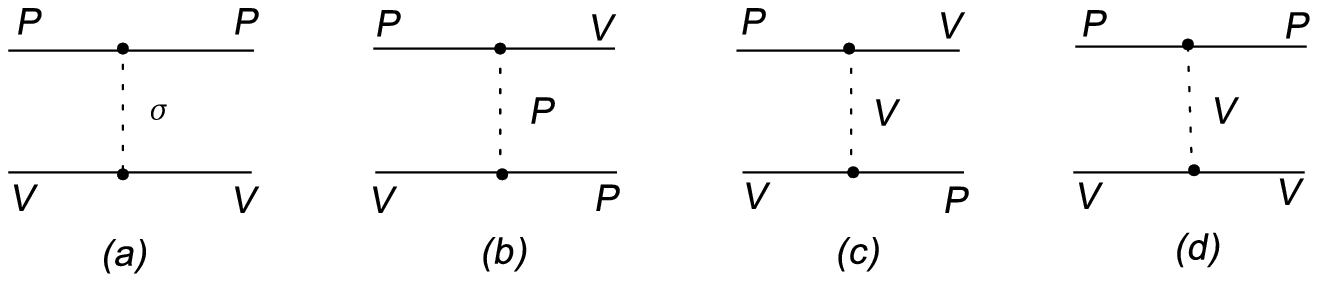}}
\end{tabular}
\caption{The diagrams in the derivation of the effective potential
of the P-V system. \label{PV}}
\end{figure}
\end{center}

Fig. \ref{PV} shows the the general scattering channels in the
derivation of the exchange potential. Different from the P-P and
V-V systems, we need consider two additional crossed scattering
diagrams for the P-V system, i.e. Fig. \ref{PV} (b) and (c).

In terms of the effective Lagrangian in Sec. \ref{sec3}, the
$\sigma$ meson exchange potential in the momentum space is
\footnote{We missed a minus sign in the sigma meson exchange
potential in Ref. \cite{liu-3872}.}
\begin{eqnarray}
\mathcal{V}_{\sigma}(\mathbf{q})&=&-\frac{g_{\sigma}^2}{\mathbf{q}^2+m_{\sigma}^2}\;
.
\end{eqnarray}
The pseudoscalar meson exchange potential reads \cite{liu-3872}
\begin{eqnarray}
\mathcal{V}_{\mathbb{P}}(\mathbf{q})\left[I_1^{\mathbb{P}\mathcal{D}_{1}\mathcal{D}_{2}^*},
I_2^{\mathbb{P}\mathcal{ D}^*_{2}\mathcal{
D}_{1}},m_{\mathcal{D}_1},m_{\mathcal{D}_2^*},m_{\mathbb{P}}\right]
&=&-\frac{1}{12m_{\mathcal{D}_1}m_{\mathcal{D}^*_2}}g^2_{\mathcal{D^*D}\mathbb{P}}\,
I_1^{\mathbb{P}\mathcal{D}_{1}\mathcal{D}_{2}^*}\,
I_2^{\mathbb{P}\mathcal{ D}^*_{2}\mathcal{ D}_{1}}\,\left\{
{\begin{array}{ccc}\frac{\mathbf{q}^2}{\mu^2-\mathbf{q}^2}\quad& \mathrm{for}\,\pi \\
                 -\frac{\mathbf{q}^2}{\mathbf{q}^2+\mu^{\prime 2}}\quad&\quad\, \mathrm{for}\,\eta \end{array}}\right.
,
\end{eqnarray}
where $\mu=\sqrt{q_{0}^2-m_{\pi}^2}$,
$\mu'=\sqrt{m_{\eta(K)}^2-q_{0}^2}$ and $q_0=m_{D_2^*}-m_{D_1}$.
For the vector meson exchange, there exist two types of potentials
from the direct and crossed scattering channels:
\begin{eqnarray}
\mathcal{V}_{\mathbb{V}}^{Direct}(\mathbf{q})\left[I_1^{\mathbb{V}\mathcal{D}_{1}\mathcal{D}_{1}},
I_2^{\mathbb{V}\mathcal{ D}^*_{2}\mathcal{
D}^{*}_{2}},m_{\mathcal{D}_1},m_{\mathcal{D}_2^*},m_{\mathbb{V}}\right]
&=&-g_{\mathcal{DD}\mathbb{V}}\,g_{\mathcal{D^*D^*}\mathbb{V}}\,
I_1^{\mathbb{V}\mathcal{D}_{1}\mathcal{D}_{1}}\,
I_2^{\mathbb{V}\mathcal{ D}^*_{2}\mathcal{
D}^{*}_{2}}\,\left(\frac{\mathbf{q}^2}{4m_{D_1}m_{D_{2}^*}m_{\mathbb{V}}^2}
+\frac{1}{\mathbf{q}^2+m_{\mathbb{V}}^2}\right),
\end{eqnarray}
\begin{eqnarray}
\mathcal{V}_{\mathbb{V}}^{Cross}(\mathbf{q})\left[I_1^{\mathbb{V}\mathcal{D}_{1}\mathcal{
D}^*_{2}}, I_2^{\mathbb{V}\mathcal{
D}^{*}_{2}\mathcal{D}_{1}},m_{\mathcal{D}_1},m_{\mathcal{D}_2^*},m_{\mathbb{V}}\right]
&=&-\frac{8}{3}f_{\mathcal{D^*D}\mathbb{V}}^2\,I_1^{\mathbb{V}\mathcal{D}_{1}\mathcal{
D}^*_{2}}\, I_2^{\mathbb{V}\mathcal{
D}^{*}_{2}\mathcal{D}_{1}}\,\frac{\mathbf{q}^2}{\mathbf{q}^2+\zeta^2},
\end{eqnarray}
where $\zeta_{\mathcal{V}}=\sqrt{m_{\mathcal{V}}^2-q_0^2}$. One
notes that $m_{\mathcal{V}}$ is larger than $q_0$ for the P-V
case. After Fourier transformation, we obtain the effective
potentials in the coordinate space
\begin{eqnarray}
\mathcal{V}_{\sigma}(r)&=&-{g_{\sigma}^2}Y[\Lambda,m_{\sigma},r],
\end{eqnarray}
\begin{eqnarray}
\mathcal{V}_{\mathbb{P}}(r)\left[I_1^{\mathbb{P}\mathcal{D}_{1}\mathcal{D}_{2}^*},
I_2^{\mathbb{P}\mathcal{ D}^*_{2}\mathcal{
D}_{1}},m_{\mathcal{D}_1},m_{\mathcal{D}_2^*},m_{\mathbb{P}}\right]
&=&-\frac{1}{12m_{\mathcal{D}_1}m_{\mathcal{D}^*_2}}g^2_{\mathcal{D^*D}\mathbb{P}}\,
I_1^{\mathbb{P}\mathcal{D}_{1}\mathcal{D}_{2}^*}\,
I_2^{\mathbb{P}\mathcal{ D}^*_{2}\mathcal{ D}_{1}}\,\left\{
{\begin{array}{ccc}U(r)\quad& \mathrm{for}\,\pi \\
                -Z[\Lambda,\mu',r]\quad&\quad\, \mathrm{for}\,\eta \end{array}}\right.
,
\end{eqnarray}
\begin{eqnarray}
\mathcal{V}_{\mathbb{V}}^{Direct}(r)\left[I_1^{\mathbb{V}\mathcal{D}_{1}\mathcal{D}_{1}},
I_2^{\mathbb{V}\mathcal{ D}^*_{2}\mathcal{
D}^{*}_{2}},m_{\mathcal{D}_1},m_{\mathcal{D}_2^*},m_{\mathbb{V}}\right]
&=&-g_{\mathcal{DD}\mathbb{V}}\,g_{\mathcal{D^*D^*}\mathbb{V}}\,
I_1^{\mathbb{V}\mathcal{D}_{1}\mathcal{D}_{1}}\,
I_2^{\mathbb{V}\mathcal{ D}^*_{2}\mathcal{
D}^{*}_{2}}\,\left(\frac{X[\Lambda,m_{\mathbb{V}},r]}{{4m_{D_1}m_{D_{2}^*}m_{\mathbb{V}}^2}}
+Y[\Lambda,m_{\mathbb{V}},r]\right),\nonumber\\
\end{eqnarray}
\begin{eqnarray}
\mathcal{V}_{\mathbb{V}}^{Cross}(r)\left[I_1^{\mathbb{V}\mathcal{D}_{1}\mathcal{
D}^*_{2}}, I_2^{\mathbb{V}\mathcal{
D}^{*}_{2}\mathcal{D}_{1}},m_{\mathcal{D}_1},m_{\mathcal{D}_2^*},m_{\mathbb{V}}\right]
&=&-\frac{8}{3}f_{\mathcal{D^*D}\mathbb{V}}^2\,I_1^{\mathbb{V}\mathcal{D}_{1}\mathcal{
D}^*_{2}}\, I_2^{\mathbb{V}\mathcal{
D}^{*}_{2}\mathcal{D}_{1}}\,Z[\Lambda,\zeta_{\mathcal{V}},r].
\end{eqnarray}
Here $$U(r)=-\frac{\mu^2}{4\pi r}\left[\cos(\mu r)-e^{-\alpha
r}\right]-\frac{\beta^2\alpha}{8\pi}e^{-\alpha r}$$ with
$\beta=\sqrt{\Lambda^2-m_{\pi}^2}$ and
$\alpha=\sqrt{\Lambda^2-q_0^2}$. The total effective potentials of
$\Phi_{s}^{*\pm}/\hat \Phi_{s}^{*\pm},\Phi_s^{*0}/\hat
\Phi_s^{*0},\bar\Phi_{s}^{*0}/\hat{\bar{\Phi}}_{s}^{*0}$
\begin{eqnarray}
\mathcal{V}(r)_{Total}^{\Phi_s^{*\pm}/\hat
\Phi_s^{*\pm}}&=&c\,\frac{g_{\mathcal{D^*D}\mathbb{P}}^2}
{36\sqrt{m_{D}m_{D_s}m_{D^*}m_{D_s^*}}}Z[\Lambda,\mu',r].
\end{eqnarray}
For $\Phi^{*\pm}/\hat\Phi^{*\pm}$, the total effective potential
reads
\begin{eqnarray}
\mathcal{V}(r)_{Total}^{\Phi^{*\pm}/\hat
\Phi^{*\pm}}&=&\frac{1}{2}g_{\mathcal{DD}\mathbb{V}}g_{\mathcal{D^*D^*}\mathbb{V}}
\left[\frac{X[\Lambda,m_{\rho},r]}{4m_{D}m_{D^*}m_{\rho}^2}+Y[\Lambda,m_{\rho},r]\right]
-\frac{1}{2}g_{\mathcal{DD}\mathbb{V}}g_{\mathcal{D^*D^*}\mathbb{V}}
\left[\frac{X[\Lambda,m_{\omega},r]}{4m_{D}m_{D^*}m_{\omega}^2}+Y[\Lambda,m_{\omega},r]\right]\nonumber\\&&
-c\,\left\{-\frac{g^2_{\mathcal{D^*D}\mathbb{P}}}{24m_{D}m_{D^*}}\,U(r)\,
-\frac{g^2_{\mathcal{D^*D}\mathbb{P}}}{72m_{D}m_{D^*}}Z[\Lambda,\mu',r]-\frac{4}{3}f^2_{\mathcal{D^*D}\mathbb{V}}
Z[\Lambda,\zeta_{\rho},r]+\frac{4}{3}f^2_{\mathcal{D^*D}\mathbb{V}}
Z[\Lambda,\zeta_{\omega},r]
\right\}\nonumber\\&&-{g_{\sigma}^2}Y[\Lambda,m_{\sigma},r].
\end{eqnarray}
For $\Phi^{*0}/\hat{\Phi}^{*0}$,
$\Phi_{s1}^{*0}/\hat{\Phi}_{s1}^{*0}$ and
$\Phi_{8}^{*0}/\hat{\Phi}_8^{*0}$, we have
\begin{eqnarray}
\mathcal{V}(r)_{Total}^{\Phi^{*0}/\hat{\Phi}^{*0}}&=&\frac{1}{2}g_{\mathcal{DD}\mathbb{V}}g_{\mathcal{D^*D^*}\mathbb{V}}
\left[\frac{X[\Lambda,m_{\rho},r]}{4m_{D}m_{D^*}m_{\rho}^2}+Y[\Lambda,m_{\rho},r]\right]
-\frac{1}{2}g_{\mathcal{DD}\mathbb{V}}g_{\mathcal{D^*D^*}\mathbb{V}}
\left[\frac{X[\Lambda,m_{\omega},r]}{4m_{D}m_{D^*}m_{\omega}^2}+Y[\Lambda,m_{\omega},r]\right]\nonumber\\&&
-c\,\left\{-\frac{g^2_{\mathcal{D^*D}\mathbb{P}}}{24m_{D}m_{D^*}}\,U(r)\,
-\frac{g^2_{\mathcal{D^*D}\mathbb{P}}}{72m_{D}m_{D^*}}Z[\Lambda,\mu',r]-\frac{4}{3}f^2_{\mathcal{D^*D}\mathbb{V}}
Z[\Lambda,\zeta_{\rho},r]+\frac{4}{3}f^2_{\mathcal{D^*D}\mathbb{V}}
Z[\Lambda,\zeta_{\omega},r]
\right\}\nonumber\\&&-{g_{\sigma}^2}Y[\Lambda,m_{\sigma},r],
\end{eqnarray}
\begin{eqnarray}
\mathcal{V}(r)_{Total}^{\Phi_{s1}^{*0}/\hat{\Phi}_{s1}^{*0}}&=&-g_{\mathcal{DD}\mathbb{V}}g_{\mathcal{D^*D^*}\mathbb{V}}
\left[\frac{X[\Lambda,m_{\phi},r]}{4m_{D_s}m_{D_{s}^*}m_{\phi}^2}+Y[\Lambda,m_{\phi},r]\right]
-c\,\frac{8}{3}f_{D^*D\mathcal{V}}^2Z[\Lambda,\zeta_{\phi},r]+c\,\frac{g_{\mathcal{D^*D}\mathbb{P}}^2}
{18{m_{D_s}m_{D_s^*}}}Z[\Lambda,\mu',r],\nonumber\\
\end{eqnarray}
\begin{eqnarray}
\mathcal{V}(r)_{Total}^{\Phi_8^{*0}/\hat{\Phi}_8^{*0}}&=&-\frac{3}{2}g_{\mathcal{DD}\mathbb{V}}g_{\mathcal{D^*D^*}\mathbb{V}}
\left[\frac{X[\Lambda,m_{\rho},r]}{4m_{D}m_{D^*}m_{\rho}^2}+Y[\Lambda,m_{\rho},r]\right]
-\frac{1}{2}g_{\mathcal{DD}\mathbb{V}}g_{\mathcal{D^*D^*}\mathbb{V}}
\left[\frac{X[\Lambda,m_{\omega},r]}{4m_{D}m_{D^*}m_{\omega}^2}+Y[\Lambda,m_{\omega},r]\right]\nonumber\\&&
-\,c\,\left\{\frac{g^2_{\mathcal{D^*D}\mathbb{P}}}{8m_{D}m_{D^*}}\,U(r)\,
-\frac{g^2_{\mathcal{D^*D}\mathbb{P}}}{72m_{D}m_{D^*}}Z[\Lambda,\mu',r]+4f^2_{\mathcal{D^*D}\mathbb{V}}
Z[\Lambda,\zeta_{\rho},r]+\frac{4}{3}f^2_{\mathcal{D^*D}\mathbb{V}}
Z[\Lambda,\zeta_{\omega},r]
\right\}\nonumber\\&&-{g_{\sigma}^2}Y[\Lambda,m_{\sigma},r]
\end{eqnarray}
The parameter $c$ is  $\mp1$ for the P-V systems with the positive
and negative charge parity respectively. As expected, the
potential of $\Phi^{*0}$ and $\hat{\Phi}^{*\pm}$ are the same as
that of $\Phi^{*\pm}$ and $\hat{\Phi}^{*0}$ respectively.

\subsection{Numerical results for the P-V system} \label{result-pv}

\begin{figure}[htb]
\begin{center}
\begin{tabular}{ccc}
\scalebox{0.55}{\includegraphics{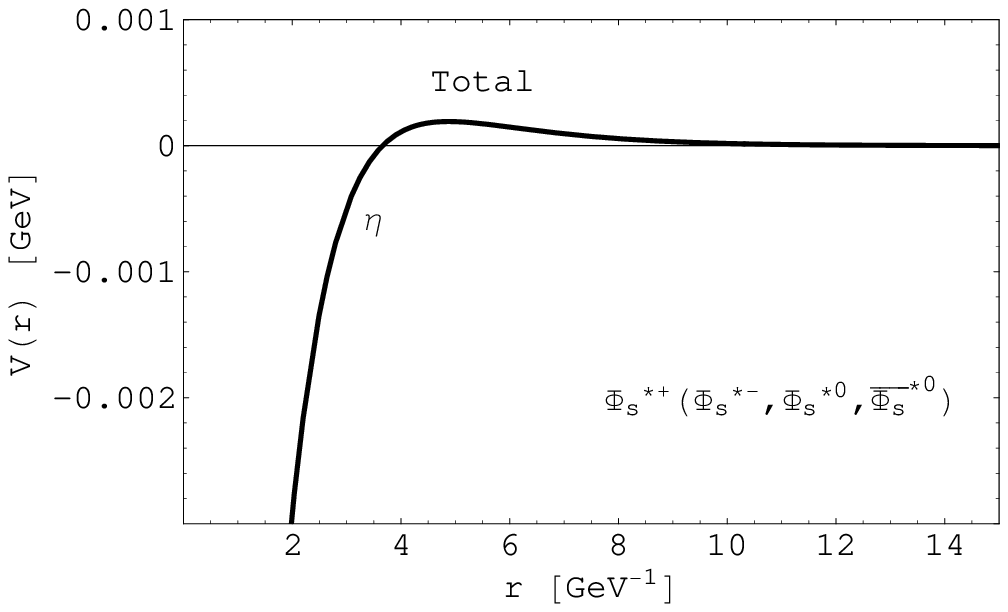}}&
\scalebox{0.55}{\includegraphics{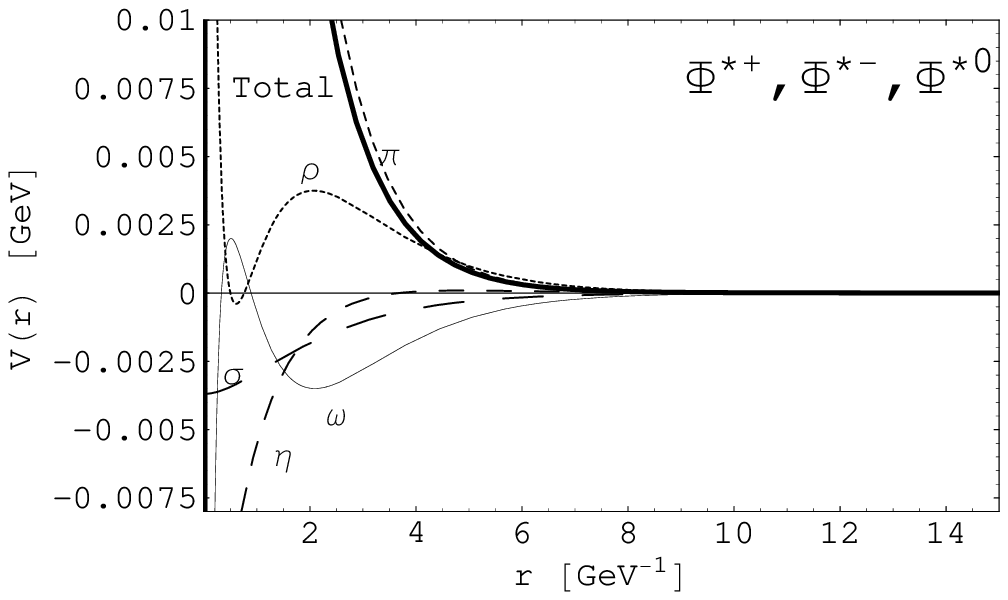}}&
\scalebox{0.55}{\includegraphics{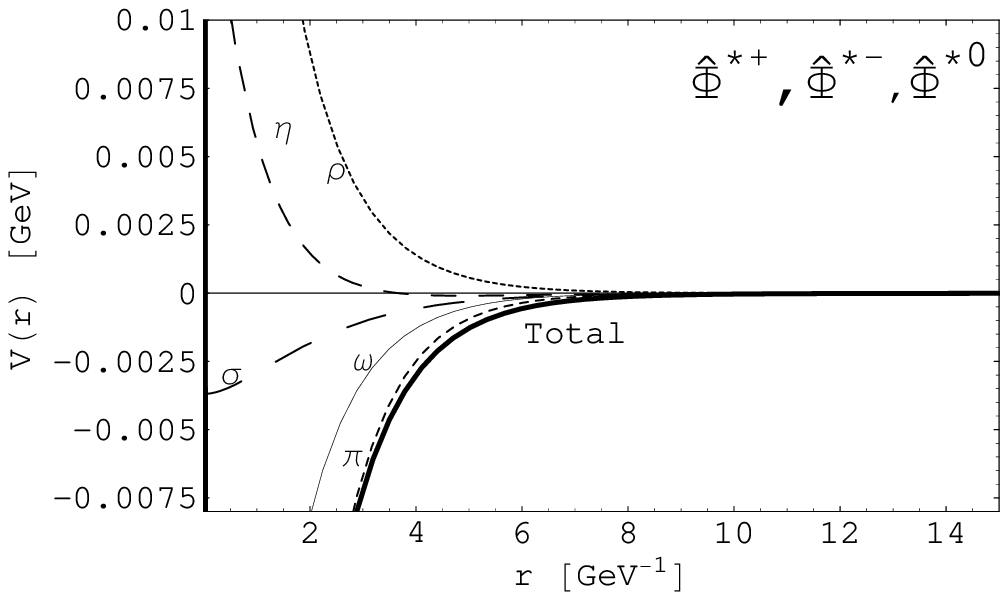}}\\(a)&(b)&(c)\\
\scalebox{0.55}{\includegraphics{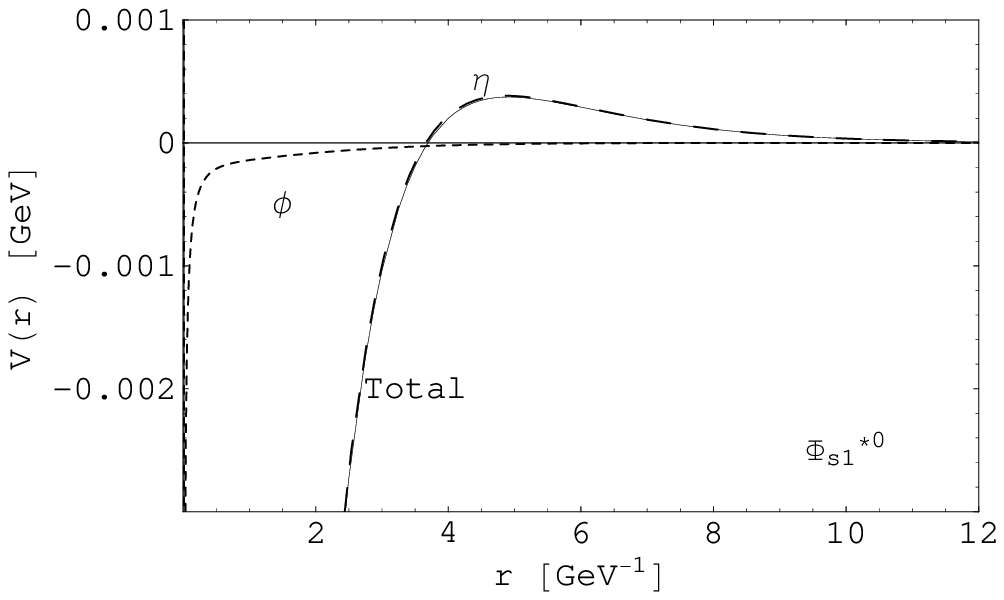}}&
\scalebox{0.55}{\includegraphics{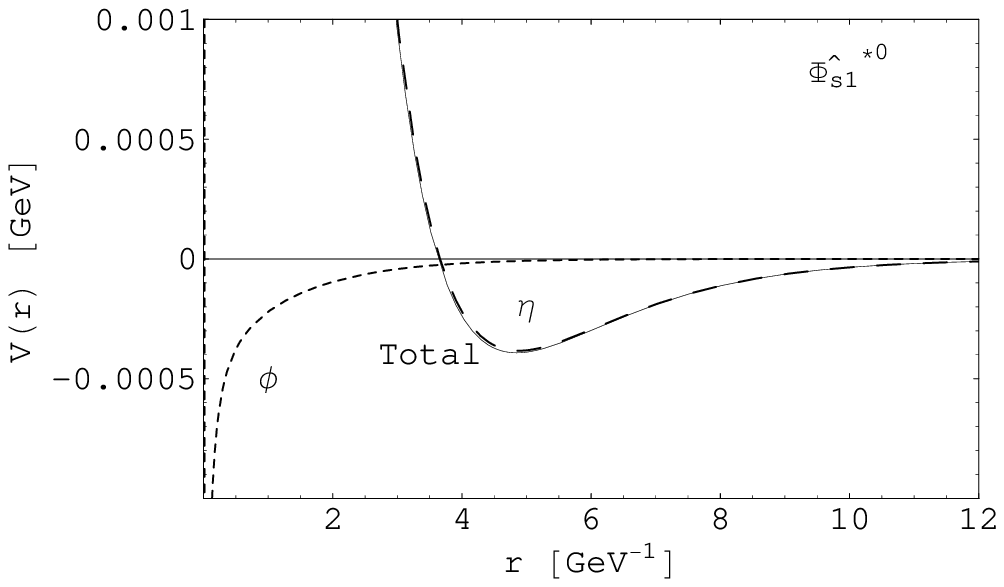}}&
\scalebox{0.55}{\includegraphics{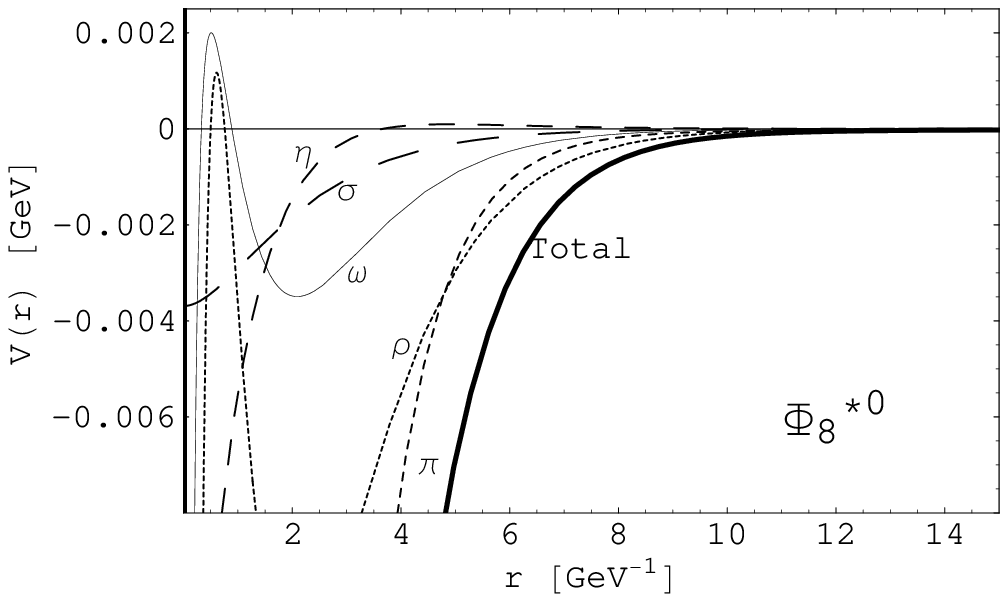}}\\(d)&(e)&(f)\\
\scalebox{0.55}{\includegraphics{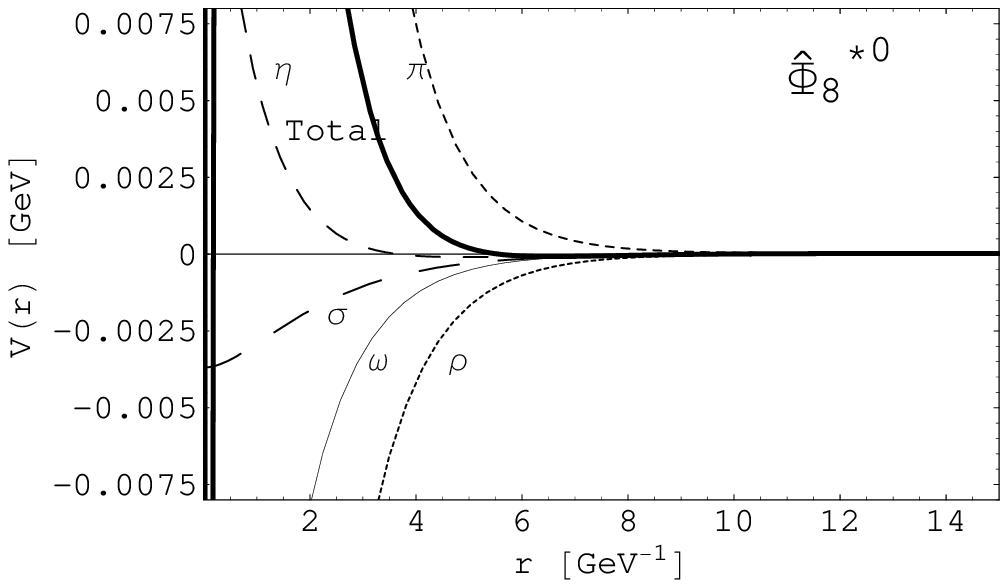}}&&\\(g)&&\\
\end{tabular}
\caption{The exchange potential of the P-V system with the typical
value $\Lambda=1$ GeV. The thick solid line is the total effective
potential. \label{PV-TYPE}}
\end{center}
\end{figure}

\begin{center}
\begin{figure}[htb]
\begin{tabular}{c}
\scalebox{0.6}{ \includegraphics{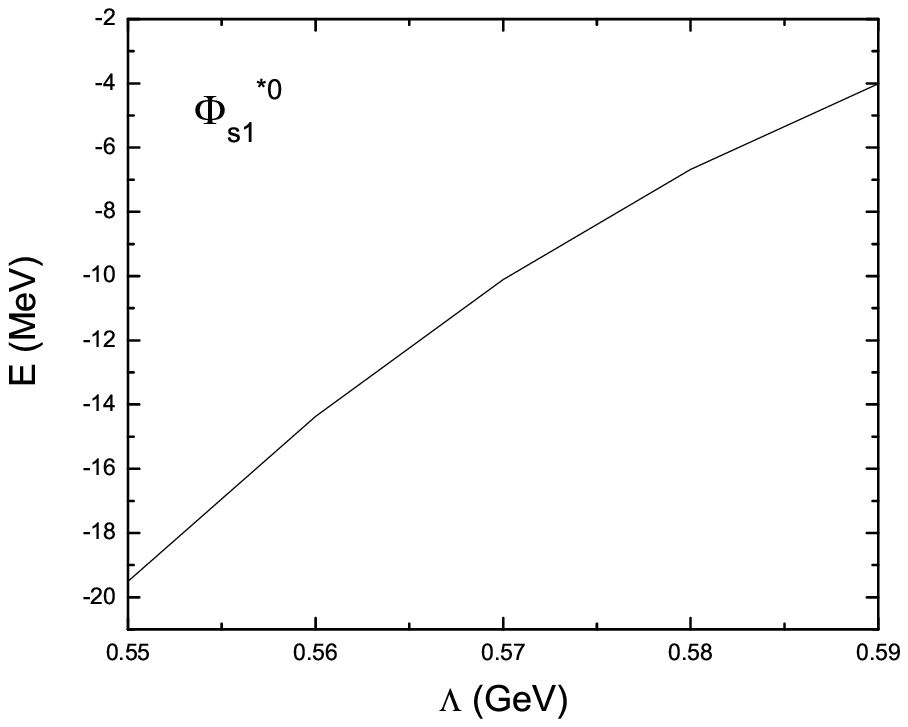}}
\end{tabular} \caption{The dependence of $E$ on $\Lambda$ for $\Phi_{s1}^{*0}$. \label{PV-Phi_s1}}
\end{figure}
\end{center}

We show the variation of the potential with $r$ in Fig.
\ref{PV-TYPE}. The total effective potential for
$\hat\Phi^{*\pm}(\hat\Phi^{*0})$ and $\Phi_{8}^{*0}$ is attractive
while the potential of $\Phi^{*\pm}({\Phi}^{*0})$ and
$\hat{\Phi}_8^{*0}$ is repulsive with $\Lambda=1$ GeV. Especially
the neutral states with positive charge parity are very
interesting. The only candidate of $X(3872)$ with the attractive
potential is $\Phi_{8}^{*0}$. We collect the numerical results in
Table \ref{DDstar}. There do not exist molecular states
$\Phi^{\*\pm}(\Phi^{*0})$, $\Omega^{\*\pm}(\Omega^{*0})$ and
${\hat \Omega}_8^{*0}$ because of the strong repulsive potential.
Bound state solutions exist for
$\Phi_{s}^{*\pm}(\Phi_{s}^{*0},\bar{\Phi}_{s}^{*0})$ and
$\Omega_{s}^{*\pm}(\Omega_{s}^{*0},\bar\Omega_{s}^{*0})$,
$\hat\Omega^{*\pm}(\hat\Omega^{*0})$ only with a very large
$\Lambda$. We show the dependence of the binding energy of
$\Phi_{s1}^0$ on $\Lambda$ in Fig. \ref{PV-Phi_s1}. Moreover we
collect the numerical results in Table \ref{DDstar-2times} when
all coupling constants are increased by a factor of two. When all
the coupling constants are reduced by a factor of two, the results
are shown in Table \ref{DDstar-half}.

\begin{center}
\begin{ruledtabular}
\begin{table}[htb]
\begin{tabular}{c||ccccccccc}
&\multicolumn{3}{c}{$\mathcal{D-\bar D^*}$} \\\hline
State&$\Lambda$ (GeV)&$E$ (MeV)& $r_{\mathrm{rms}}$ (fm)\\\hline

$\hat\Phi_s^{*\pm},\hat\Phi_{s}^{*0},\hat\bar{\Phi}_{s}^{*0}$&-&-&-\\\hline

$\Phi_s^{*\pm},\Phi_{s}^{*0},\bar{\Phi}_{s}^{*0}$&8.90&-11.74&1.29\\
                                                 &9.00&-25.38&0.88\\ \hline

$\hat\Phi^{*\pm},\hat\Phi^{*0}$&10.5&-12.86&1.35\\\hline

${\Phi}^{*\pm}, {\Phi}^{*0} $&-&-&-\\\hline

$\Phi_{s1}^{*0}$&0.55&-19.51&1.55\\
                &0.57&-10.11&1.93\\\hline

$\hat\Phi_{s1}^{*0}$&0.60&-17.96&1.43\\
                    &0.62&-6.94&2.01\\\hline

$\Phi_{8}^{*0}$&0.52&-13.02&1.79\\
               &0.54&-5.25&2.38\\\hline

$\hat\Phi_{8}^{*0}$&0.50&-28.61&1.30\\
                   &0.52&-9.42&1.91\\\hline

&\multicolumn{3}{c}{$\mathcal{B-\bar B}^*$} \\\hline
State&$\Lambda$ (GeV)&$E$ (MeV)& $r_{\mathrm{rms}}$ (fm)\\\hline
$\hat\Omega_{s}^{*\pm},\hat\Omega_{s}^{*0},\hat{\bar\Omega}_{s}^{*0}$&-&-&-\\\hline

$\Omega_{s}^{*\pm},\Omega_{s}^{*0},\bar\Omega_{s}^{*0}$&3.80&-3.07&2.20\\
                                                       &3.90&-20.23&1.00\\\hline

$\hat\Omega^{*\pm},\hat\Omega^{*0}$&2.80&-4.20&1.41\\
                               &3.00&-21.13&0.96\\\hline

$\Omega^{*\pm},\Omega^{*0}$&-&-&-\\\hline

$\Omega_{s1}^{*0}$& 2.10&-5.94&1.99\\
                  &2.20 &-9.79&1.65\\\hline

$\hat\Omega_{s1}^{*0}$&0.72&-12.40&1.59\\
                      &0.73&-5.95&2.08\\\hline

$\Omega_{8}^{*0}$&0.92&-10.19&1.13\\
                 &0.94&-18.67&1.03\\\hline

$\hat\Omega_{8}^{*0}$&-&-&-\\

                       \end{tabular}
\caption{The possible bound state solutions for the P-V system.
\label{DDstar}}
\end{table}
\end{ruledtabular}
\end{center}

\begin{center}
\begin{ruledtabular}
\begin{table}[htb]
\begin{tabular}{c||ccccccccc}
&\multicolumn{3}{c}{$\mathcal{D-\bar D^*}$} \\\hline
State&$\Lambda$ (GeV)&$E$ (MeV)& $r_{\mathrm{rms}}$ (fm)\\\hline

$\hat\Phi_s^{*\pm},\hat\Phi_{s}^{*0},\hat{\bar{\Phi}}_{s}^{*0}$&-&-&-\\\hline

$\Phi_s^{*\pm},\Phi_{s}^{*0},\bar{\Phi}_{s}^{*0}$&2.80&-1.58&2.57\\
                                                 &2.90&-15.38&1.15\\ \hline

$\hat\Phi^{*\pm},\hat\Phi^{*0}$&1.90&-16.33&1.27\\
                            &2.00&--26.25&1.03\\\hline

${\Phi}^{*\pm}, {\Phi}^{*0} $&-&-&-\\\hline

$\Phi_{s1}^{*0}$&0.72&-13.66&1.65\\
                &0.73&-8.65&1.93\\\hline

$\hat\Phi_{s1}^{*0}$&0.75&-22.85&1.24\\
                    &0.77&-4.66&2.24\\\hline

$\Phi_{8}^{*0}$&0.75&-28.48&1.15\\
               &0.78&-35.20&1.05\\\hline

$\hat\Phi_{8}^{*0}$&1.21&-4.47&2.29\\
                   &1.23&-19.31&1.26\\\hline

&\multicolumn{3}{c}{$\mathcal{B-\bar B}^*$} \\\hline
State&$\Lambda$ (GeV)&$E$ (MeV)& $r_{\mathrm{rms}}$ (fm)\\\hline

$\hat\Omega_s^{*\pm},\hat\Omega_{s}^{*0},\hat{\bar{\Omega}}_{s}^{*0}$&-&-&-\\\hline

$\Omega_{s}^{*\pm},\Omega_{s}^{*0},\bar\Omega_{s}^{*0}$&1.60&-6.24&1.74\\
                                                       &1.70&-44.08&0.74\\\hline

$\hat\Omega^{*\pm},\hat\Omega^{*0}$&0.87&-5.65&1.12\\
                               &0.90&-12.67&1.03\\\hline

$\Omega^{*\pm},\Omega^{*0}$&-&-&-\\\hline

$\Omega_{s1}^{*0}$& 1.15&-5.70&1.88\\
                  &1.20&-30.52&0.96\\\hline

$\hat\Omega_{s1}^{*0}$&0.82&-34.13&1.07\\
                      &0.84&-3.57&2.43\\\hline

$\hat\Omega_{8}^{*0}$&-&-&-\\

                       \end{tabular}
\caption{The possible bound state solutions for the P-V system
after increasing all the coupling constants by a factor of two.
\label{DDstar-2times}}
\end{table}
\end{ruledtabular}
\end{center}

\begin{center}
\begin{ruledtabular}
\begin{table}[htb]
\begin{tabular}{c||ccccccccc}
&\multicolumn{3}{c}{$\mathcal{D-\bar D^*}$} \\\hline
State&$\Lambda$ (GeV)&$E$ (MeV)& $r_{\mathrm{rms}}$ (fm)\\\hline

$\hat\Phi_s^{*\pm},\hat\Phi_{s}^{*0},\hat{\bar{\Phi}}_{s}^{*0}$&-&-&-\\\hline

$\Phi_s^{*\pm},\Phi_{s}^{*0},\bar{\Phi}_{s}^{*0}$&-&-&-\\ \hline

$\hat\Phi^{*\pm},\hat\Phi^{*0}$&-&-&-\\ \hline

 ${\Phi}^{*\pm},
{\Phi}^{*0} $&0.36&-11.10&2.07\\
               &0.37&-7.30&2.38 \\\hline

$\Phi_{s1}^{*0}$&0.39&-17.22&1.74\\
                &0.41&-9.31&2.11\\\hline

$\hat\Phi_{s1}^{*0}$&0.43&-12.33&1.78\\
                    &0.45&-5.42&2.32\\\hline

$\Phi_{8}^{*0}$&0.44&-18.01&1.68\\
               &0.46&-8.66&2.11\\\hline

$\hat\Phi_{8}^{*0}$&0.38&-13.56&1.90\\
&0.40&-6.17&2.41\\                   \hline

&\multicolumn{3}{c}{$\mathcal{B-\bar B}^*$} \\\hline
State&$\Lambda$ (GeV)&$E$ (MeV)& $r_{\mathrm{rms}}$ (fm)\\\hline

$\hat\Omega_s^{*\pm},\hat\Omega_{s}^{*0},\hat{\bar{\Omega}}_{s}^{*0}$&-&-&-\\\hline

$\Omega_{s}^{*\pm},\Omega_{s}^{*0},\bar\Omega_{s}^{*0}$&-&-&\\\hline

$\hat\Omega^{*\pm},\hat\Omega^{*0}$&-&-&\\\hline

$\Omega^{*\pm},\Omega^{*0}$&-&-&-\\\hline

$\Omega_{s1}^{*0}$& -&-&\\\hline

$\hat\Omega_{s1}^{*0}$&3.30&-1.39&2.71\\
                      &3.40&-11.03&1.41\\\hline

$\Omega_{8}^{*0}$&0.46&-19.73&1.50\\
                 &0.48&-8.10&1.78\\\hline

$\hat\Omega_{8}^{*0}$&-&-&-\\

                       \end{tabular}
\caption{The possible bound state solutions for the P-V system
after reducing all the coupling constants by a factor of two.
\label{DDstar-half}}
\end{table}
\end{ruledtabular}
\end{center}

\section{Discussion and conclusion}\label{sec5}

We have systematically studied the P-P, V-V and P-V systems, which
is composed of a pair of heavy meson and anti-meson. We summarize
our numerical results from Secs. \ref{secpp}-\ref{secpv} in Tables
\ref{remark-PP}, \ref{remark-VV} and \ref{remark-PV}. We use the
symbols '$\bigcirc$' and '$\bigotimes$' to indicate whether a
molecular state exists or not with the coupling constants in
Section \ref{sec3}. In the following two cases, we tend to label
the system with the symbol '$\bigotimes$': (1) there does not
exist a bound state at all because of the repulsive potential; (2)
a bound state solution exists only with either $\Lambda >3$ GeV or
$\Lambda <0.5$ GeV. If a bound state exists when $0.5< \Lambda
<0.9$ GeV or $2.0< \Lambda <3.0$ GeV, we mark this state with '?'.
In this case the result is so sensitive to the cutoff parameter
that we can not draw a definite conclusion. Only when a loosely
molecular state exists with $0.9< \Lambda <2.0$ GeV, we tend to
label the state with the symbol '$\bigcirc$'. We want to emphasize
the above criteria may be biased due to the authors' personal
judgement. The readers are encouraged to consult the numerical
results in Sections \ref{secpp}-\ref{secpv} and draw their own
conclusions.

\begin{center}
\begin{table}[htb]
\begin{tabular}{c|c||c|cccccc}\hline\hline
\multicolumn{2}{c}{$\mathcal{D-\bar D}$}
&\multicolumn{2}{c}{$\mathcal{B-\bar B}$} \\\hline

state& remark & state & remark \\\hline

$\Phi_{s}^{\pm}, \Phi_{s}^{0}, \bar \Phi_{s}^0$& $\bigotimes$
&$\Omega_{s}^{\pm}, \Omega_{s}^{0}, \bar \Omega_{s}^0$&
$\bigotimes$\\

$\Phi^{\pm}, \Phi^{0}$& $\bigotimes$ &$\Omega^{\pm}, \Omega^{0}$&
$\bigotimes$\\

$\Phi_8^{0}$& $?$ &$\Omega_8^{0}$&
$\bigcirc$\\

$\Phi_{s1}^{0}$& $?$ &$\Omega_{s1}^{0}$& $?$\\\hline\hline

            \end{tabular}
\caption{Summary of the $\mathcal{D-\bar D}$ and $\mathcal{B-\bar
B}$ systems. The symbols $\bigotimes$ and $\bigcirc$ denote
"forbidden" and "allowed" respectively. \label{remark-PP}}
\end{table}
\end{center}

\begin{center}
\begin{table}[htb]
\begin{tabular}{c|ccc||c|ccccc}\hline\hline
\multicolumn{4}{c}{$\mathcal{D^*-\bar D^*}$}
&\multicolumn{4}{c}{$\mathcal{B^*-\bar B^*}$} \\\hline

state& \multicolumn{3}{c}{remark} & state &
\multicolumn{3}{c}{remark} \\\hline

&$J^{P}=0^+$&$J^{P}=1^+$&$J^{P}=2^+$&&$J^{P}=0^+$&$J^{P}=1^+$&$J^{P}=2^+$\\\hline

$\Phi_{s}^{**\pm}, \Phi_{s}^{**0}, \bar \Phi_{s}^{**0}$&
$\bigotimes$& $\bigotimes$&$\bigotimes$&$\Omega_{s}^{**\pm},
\Omega_{s}^{**0}, \bar
\Omega_{s}^{**0}$& $?$&$\bigotimes$&$\bigotimes$\\

$\Phi^{**\pm}, \Phi^{**0}$& $\bigotimes$&
$\bigotimes$&$\bigotimes$&$\Omega^{**\pm}, \Omega^{**0}$& $\bigcirc$&$?$&$\bigotimes$\\

$\Phi_8^{**0}$& $?$&
$?$&$?$&$\Omega_8^{**0}$& $?$&$?$&$?$\\

$\Phi_{s1}^{**0}$& $?$& $?$&$?$&$\Omega_{s1}^{**0}$&
$?$&$?$&$?$\\\hline\hline
            \end{tabular}
\caption{Summary of the $\mathcal{D^*-\bar D^*}$ and
$\mathcal{B^*-\bar B^*}$ systems. \label{remark-VV}}
\end{table}
\end{center}

\begin{center}
\begin{table}[htb]
\begin{tabular}{c|c||c|cccccc}\hline\hline
\multicolumn{2}{c}{$\mathcal{D-\bar D^*}$}
&\multicolumn{2}{c}{$\mathcal{B-\bar B^*}$} \\\hline

state& remark & state & remark \\\hline

$\Phi_{s}^{*\pm}, \Phi_{s}^{*0}, \bar \Phi_{s}^{*0}$& $\bigotimes$
&$\Omega_{s}^{*\pm}, \Omega_{s}^{*0}, \bar \Omega_{s}^{*0}$&
$\bigotimes$\\

$\hat\Phi_s^{*\pm},\hat\Phi_{s}^{*0},\hat{\bar{\Phi}}_{s}^{*0}$&
$\bigotimes$
&$\hat\Omega_s^{*\pm},\hat\Omega_{s}^{*0},\hat{\bar{\Omega}}_{s}^{*0}$&$\bigotimes$\\

$\Phi^{*\pm}, \Phi^{*0}$& $\bigotimes$ &$\Omega^{*\pm},
\Omega^{*0}$&
$\bigotimes$\\

$\hat\Phi^{*\pm},\hat\Phi^{*0}$&$\bigotimes$&$\hat\Omega^{*\pm},
\hat\Omega^{*0}$&$\bigotimes$\\

$\Phi_{s1}^{*0}$&$?$&$\Omega_{s1}^{*0}$&$?$\\

$\hat\Phi_{s1}^{*0}$&$?$&$\hat\Omega_{s1}^{*0}$&$?$\\

$\Phi_8^{*0}$& $?$ &$\Omega_8^{*0}$&
$\bigcirc$\\

$\hat\Phi_{8}^{*0}$& $?$ &$\hat\Omega_{8}^{*0}$&
$\bigotimes$\\\hline\hline

            \end{tabular}
\caption{Summary of the $\mathcal{D-\bar D^*}$ and
$\mathcal{B-\bar B^*}$ systems. \label{remark-PV}}
\end{table}
\end{center}

It's interesting to compare our results with those near-threshold
structures observed recently.
\begin{enumerate}

\item{$Z^+(4051)$\\

The charged broad structure $Z^+(4051)$ was observed by Belle in
the $\pi^+\chi_{c1}$. Its central mass value is slightly above the
$D^*\bar D^*$ threshold. First we want to emphasize that there
exists strong attraction in the range $r<1$ fm for the $D^*\bar
D^*$ system with $J=0, 1$ as can be seen clearly from the shape of
the total effective potential in Fig. \ref{VV-potential-2}
(a)-(b).

If this enhancement is a molecular state, the potential candidate
of $Z^+(4051)$ is the $\Phi^{**+}$ state with isospin $I=1$. Our
numerical results indicate that there exists a bound state
solution (1) for $\Phi^{**+}$ with $J^{P}=0^+$ and $\Lambda\sim 4$
GeV; (2) for $\Phi^{**+}$ with $J^{P}=1^+$ and $\Lambda\sim 10$
GeV. Unfortunately such a cutoff seems too large according to our
criteria.

If future experiments confirm $Z^+(4051)$ as a loosely bound
molecular state, its quantum number is very probably $J^{P}=0^+$,
which can be measured experimentally through the angular momentum
distribution of $\pi^+\chi_{c1}$. Since $\Phi^{**+}$ belongs to an
isospin triplet, its neutral partner state $\Phi^{**0}$ may be
searched in the $\pi^0\chi_{c1}$ channel.}

\item{$X(3872)$\\

In our previous work \cite{liu-3872}, we explored the $D^0 \bar
D^{*0}$ molecular state by considering the $\pi$ and $\sigma$
meson exchange potentials only. Our numerical results showed that
it is impossible to form a $D^0 \bar D^{*0}$ molecular state with
the coupling constants determined by experiment and a reasonable
cutoff around 1 GeV \cite{liu-3872}. Unfortunately we missed a
minus sign in the sigma meson exchange potential, although its
contribution is not large. Moreover we didn't take into account
the charged modes.

In this work, we consider both the neutral and charged modes and
include the exchange force from the $\pi$, $\eta$, $\sigma$,
$\rho$ and $\omega$ mesons. The resulting total effective
potential is attractive as shown in Fig. \ref{PV-TYPE} (f). The
vector meson exchange especially the rho meson exchange provides
the additional strong attraction. The $\Phi_{8}^{*0}$ state
corresponds the $X(3872)$.

\begin{center}
\begin{table}[htb]
\begin{tabular}{c|cccc}\hline\hline
$\Lambda$& $E$ (MeV)&$r_{rms}$ (fm)\\\hline

1.6&-12.86&1.47\\
1.5&-6.07&1.98\\
1.42&-2.41&3.09\\
1.41&-2.07&3.30\\
1.40&-1.75&3.53\\\hline\hline

\end{tabular}
\caption{The bound state solutions for the $D^\ast-\bar D$ system
if we fix the pionic and scalar coupling constants and reduce only
the vector coupling constant in Section \ref{sec3} by a factor of
two. \label{3872-detail}}
\end{table}
\end{center}

Recall that the sigma meson exchange force is weak and the pion
meson exchange alone does not bind the $D^{\ast}\bar D$ into a
molecular state with a physical pionic coupling constant and a
reasonable cutoff \cite{liu-3872}. In fact a bound state appears
only with $\Lambda \sim 6$ GeV there. With the additional strong
attraction from the vector meson exchange, our present numerical
analysis shows that a molecular state exists with $\Lambda\sim
0.55$ GeV. If we arbitrarily increase all the coupling constants
by a factor of two, there exists a molecular state with a slightly
more reasonable $\Lambda\sim 0.75$ GeV. If we reduce all the
coupling constants by a factor two, there appears a bound state
with $\Lambda\sim 0.45$ GeV. Such a cutoff value is below 1 GeV
and smaller than the exchanged vector meson mass.

The pionic coupling constant of the $D^\ast\bar D$ system was
extracted from the decay width of the $D^\ast$ meson
experimentally. Its value is known quite reliably. In contrast,
the vector coupling constant was estimated using the vector
dominance model in Ref. \cite{Isola}, which may be overestimated
according to a recent light cone QCD sum rule analysis \cite{lzh}.
If we fix the pionic and scalar coupling constants but reduce the
vector coupling by a factor of two only, there exists a very
reasonable bound state solution with $E=-1.75$ GeV and
$\Lambda=1.4 $GeV as shown in Table \ref{3872-detail}. Now the
value of its root-mean-square radius reaches 3.53 fm! In other
words, such a molecular state is bound very loosely.

In short summary, $X(3872)$ may be accommodated as a molecular
state dynamically, although drawing a very definite conclusion is
very difficult especially when other available experimental
information is taken into account. In our calculation, we assumed
the SU(2) symmetry and ignored the mass difference between the
charged and neutral mesons. Moreover we ignored the possible S-D
mixing effect completely since we impose the S-wave condition in
the derivation of the potential. These approximations together
with the sensitivity of the numerical results to the cutoff
parameter should be considered in the future investigation.

\item{$X(3764)$\\

The broad structure $X(3764)$ is close to the $D\bar D$ threshold
and was observed in $e^+e^-\to \mathrm{hadrons}$ process. Its
quantum number is $J^{P}=1^{--}$. In the present case, we need
consider the P-wave $\Phi_{8}^0$ by adding the centrifugal
potential $\ell(1+\ell)/{(2\mu r^2)}$ with $\ell=1$ into the
potential in Eq. (\ref{ppr-3}). Here $\mu$ is the reduced mass of
the system. In Fig. \ref{pwave}, we show the variation of
$V(r)_{Total}^{\phi_8^0}+3/(2\mu r^2)$ with several typical values
of $\Lambda$. Our numerical analysis indicates that there does not
exist a P-wave $\Phi_{8}^0$ from such a potential. In other words,
the structure $X(3764)$ is not a $D\bar D$ molecular state.

\begin{figure}[htb]\begin{center}
\begin{tabular}{ccc}
\scalebox{0.8}{\includegraphics{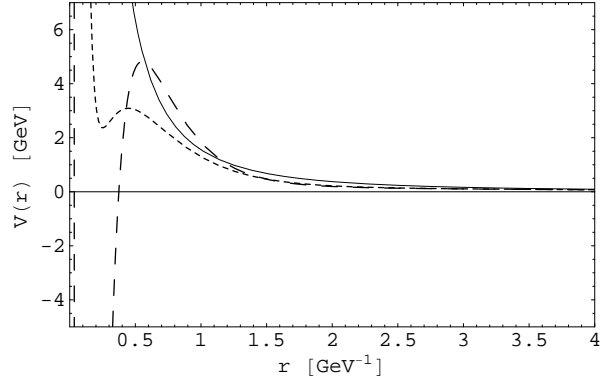}}\\
\end{tabular}
\caption{The dependence of the potential
$V(r)_{Total}^{\phi_8^0}+3/(2\mu r^2)$ on $\Lambda$. The solid,
dotted and dashed lines correspond to $\Lambda=1, 3, 4$ GeV
respectively. \label{pwave}}\end{center}
\end{figure}

}

}

\end{enumerate}

\section*{Acknowledgments}

The authors thank Professors K.T. Chao and Z.Y. Zhang for helpful
discussions. This project was supported by National Natural
Science Foundation of China under Grants 10625521, 10675008,
10705001, 10775146, 10721063 and China Postdoctoral Science
foundation (20070420526). One of authors (X.L.) thanks the support
by the \emph{Funda\c{c}\~{a}o para a Ci\^{e}ncia e a Tecnologia of
the Minist\'{e}rio da Ci\^{e}ncia, Tecnologia e Ensino Superior}
of Portugal (SFRH/BPD/34819/2007).


\end{document}